\documentclass[%
 reprint,
superscriptaddress,
nofootinbib,
 amsmath,amssymb,
 aps,
prd,
]{revtex4-2}

\usepackage{color}
\usepackage{amsmath}
\usepackage{amssymb}
\usepackage{graphicx}
\usepackage{dcolumn}
\usepackage{bm}
\usepackage{hyperref}
\usepackage{cleveref}
\usepackage{rotating}
\usepackage{pbox}
\usepackage[utf8]{inputenc}
\usepackage[]{times}
\usepackage{tikz}
\usepackage[final]{showlabels}

\usepackage{xcolor}
\definecolor{orange}{rgb}{1.0, 0.5, 0.0}

\newcommand{\rev}[1]{\textcolor{black}{#1}}

\begin{document}

\title{Electromagnetic precursor flares from the late inspiral of
  neutron star binaries}

\author{Elias R. Most}
\email{emost@princeton.edu}
\affiliation{Princeton Center for Theoretical Science, Princeton University, Princeton, NJ 08544, USA}
\affiliation{Princeton Gravity Initiative, Princeton University, Princeton, NJ 08544, USA}
\affiliation{School of Natural Sciences, Institute for Advanced Study, Princeton, NJ 08540, USA}

\author{Alexander A. Philippov}
\affiliation{Center for Computational Astrophysics, Flatiron Institute, Simons Foundation, New York, NY 10010, USA}
\affiliation{Department of Physics, University of Maryland, College Park, MD 20742, USA}

\date{\today}

\begin{abstract}
  The coalescence of two neutron stars is accompanied by the emission of
  gravitational waves, and can also feature electromagnetic counterparts powered by mass ejecta and the formation of a relativistic jet after the merger.
  Since neutron stars can feature strong magnetic fields, the non-trivial
  interaction of the neutron star magnetospheres might fuel potentially
  powerful electromagnetic transients prior to merger. 
  \rev{A key process powering} those precursor transients is relativistic reconnection
  in strong current sheets formed between the two stars.
  In this work, we provide a detailed analysis of how the twisting of the
  common magnetosphere of the binary leads to an emission of
  electromagnetic flares, akin to those produced in the solar corona.
  By means of relativistic force-free electrodynamics simulations, we
  clarify the role of different magnetic field topologies
  in the process. We conclude that flaring will always occur for suitable magnetic field alignments, unless one of
  the neutron stars has a magnetic field significantly weaker than the
  other.  
\end{abstract}


\maketitle

\section{Introduction}
\label{sec:intro}
%

With the discovery of gravitational waves from coalescing black holes
\cite{LIGOScientific:2016aoc,LIGOScientific:2016sjg,LIGOScientific:2016dsl,LIGOScientific:2017bnn,LIGOScientific:2017ycc,LIGOScientific:2018mvr,LIGOScientific:2020iuh,LIGOScientific:2020ibl,LIGOScientific:2020ufj,LIGOScientific:2020stg,LIGOScientific:2020zkf,LIGOScientific:2021djp}
(see also, e.g., \cite{Venumadhav:2019tad,Zackay:2019tzo,Olsen:2022pin}),
binary neutron star \cite{LIGOScientific:2017vwq,LIGOScientific:2020aai}
and black hole -- neutron star
\cite{LIGOScientific:2020zkf,LIGOScientific:2021qlt,LIGOScientific:2021djp}
systems, we have entered a new era of compact object science.  Especially
the matter present in the latter two systems, has the potential to fuel
electromagnetic counterparts, which where observed for the GW170817 event(
e.g.
\cite{LIGOScientific:2017ync,Mooley:2017enz,Kasliwal:2017ngb,Hallinan:2017woc,Cowperthwaite:2017dyu,Drout:2017ijr,Chornock:2017sdf,Villar:2017wcc}).
This makes the search for electromagnetic counterparts 
an integral part of any gravitational wave detection. While the first
detection of a neutron star binary coalescence was accompanied by two
distinct counterparts \cite{LIGOScientific:2017ync}, likely originating
during and after merger, there could be yet another counterpart emitted
right before merger \cite{Hansen:2000am,Lyutikov:2018nti,Wada:2020kha}.
Such a {\it precursor} is thought to arise due to interactions \rev{of
magnetic flux tubes} in the pair-plasma filled common magnetosphere of the binary
\cite{Lyutikov:2018nti},\rev{which} requires the presence a steady magnetic
field anchored in matter.  Hence, making binary neutron star and black hole
-- neutron star collisions prime targets to detect such precursors
\cite{Callister:2019biq}.  Indeed, in the case of black hole -- neutron
star mergers, electromagnetic precursors might also be the only
accompanying transient if the neutron star is swallowed whole, see e.g.
\cite{McWilliams:2011zi,DOrazio:2013ngp,Mingarelli:2015bpo,DOrazio:2015jcb,Bransgrove:2021heo}.
As such, precursors are also potential candidates
for \rev{powerful radio signals}
(e.g., \cite{Mingarelli:2015bpo,Wang:2016dgs}).  The prospects for creating
early warning systems for electromagnetic follow-up observations of
precursor emission has recently been investigated
\cite{James:2019xca,Sachdev:2020lfd,Yu:2021vvm}, crucial for their
potential future detection \cite{Wang:2020sda}.  For the event GW170817
attempts have even been made to detect precursor emission
\cite{Callister:2019biq,Broderick:2020lcv}, see also
\cite{Stachie:2021noh,Stachie:2021uky}.  Coincident detection of precursor
emission could also be used to (further) constrain the sky localization of
the merger site\cite{Gourdji:2020rca}.

A variety of models have been proposed for the production of precursors to
compact binary mergers. These include explosive fireball models
\cite{Metzger:2016mqu}, gamma-ray flares
\cite{Tsang:2011ad,Schnittman:2017nhg}, shock-powered radio precursors
\cite{Sridhar:2020uez}, unipolar inductors \cite{Lai:2012qe,Piro:2012rq},
crustal shattering \cite{Tsang:2011ad}. \rev{Particular examples of
  reconnection-based models include Refs.
  \cite{Most:2020ami,Beloborodov:2020ylo}} \rev{, where dissipation in (transient) current sheets powers the emission.} Especially the latter class of models requires a non-linear
interaction of magnetic fields in reconnecting current sheet, whose
boundary conditions are determined by the global dynamics of the problem
\cite{Cherkis:2021vto}. This motivates the need for global numerical
simulations of the magnetospheric dynamics in a close compact binary.  To
study the precursor emission mechanism numerically different types of
simulations have been performed.  In particular, simulations in
special-relativistic force-free electrodynamics have clarified the emission
of single orbiting neutron stars \cite{Carrasco:2020sxg}, idealized
mixed binaries \cite{Carrasco:2021jja}, \rev{and magnetospheric interaction in
binary neutron star systems} \cite{Most:2020ami}.  Studies in dynamical
spacetimes have focused on understanding electromagnetic bursts produced
from single neutron stars, either in isolation \cite{Palenzuela:2012my}, or
as mimickers for post-merger remnants \cite{Lehner:2011aa}. A small number
of studies has also focused on the magnetospheric dynamics
during the collapse
\cite{Dionysopoulou:2012zv,Nathanail:2017wly,Most:2018abt} and collision of
neutron stars \cite{Nathanail:2020fkp}, where \rev{their}
magnetospheres were inactive.  Full general-relativistic simulations of
force-free electrodynamics have been conducted to better understand the
common magnetosphere prior to merger, see, e.g., for black hole binaries
\cite{Palenzuela:2009yr,Palenzuela:2009hx,Mosta:2009rr,Palenzuela:2010nf},
neutron star binaries
\cite{Palenzuela:2013hu,Palenzuela:2013kra,Ponce:2014hha}, and mixed
systems \cite{East:2021spd,Paschalidis:2013jsa}.  While several numerical
formulations of relativistic force-free electrodynamics have been proposed
(e.g.,
\cite{Komissarov:2002my,McKinney:2006sc,Palenzuela:2012my,Pfeiffer:2013wza,Paschalidis:2013gma,Carrasco:2016din,Etienne:2017jmx}),
most of the simulations mentioned above have used a variation of
independently evolving electric and magnetic fields according to the
relativistic Maxwell equations \cite{Alic:2012df,Palenzuela:2012my}. The
latter is supplemented by \rev{a suitable prescription for the electric current 
mimicking a (resistive) force-free pair plasma } (see,
e.g., Refs.  \cite{Mahlmann:2020yxn,Ripperda:2021pzt,Mahlmann:2021yws} for
a recent discussion).

In this work, we extend our initial findings \cite{Most:2020ami} for the
flaring process in the common magnetosphere of neutron star binaries in
close contact. Using special-relativistic force-free electrodynamics
simulations, we present a set of models focusing on the impact of magnetic
field topology on the emission of the flares and the viability of their
production mechanism. We clarify, in particular, that the emission of
flares is largely insensitive to magnetic field alignments and unequal
field strengths, as long as there is a significant component of the
magnetic \rev{moment} pointing in each hemisphere.

This paper is structures as follows: In Sec. \ref{sec:methods} we provide a
detailed overview of the numerical methods and analysis tools used in this
work.  In our main Sec. \ref{sec:results} we provide a detailed description
of the numerical simulations presented in this work. We conclude with a
short discussion in Sec. \ref{sec:conclusions}.\\ Unless otherwise
indicated, we use geometrical units $c=G=k_B=1$.

\section{Methods}
\label{sec:methods}

We study the production of electromagnetic flares launched from an orbiting
neutron star binary in close contact. We assume that the induced electric
field at the surface of either neutron star are sufficiently large to
trigger \rev{an} electron positron pair cascade \cite{Goldreich:1969sb}, hence,
filling the immediate surrounding of the binary with a force-free highly
conductive plasma \cite{Hansen:2000am,Lyutikov:2018nti,Wada:2020kha}.
This allows us to model the binary using relativistic force-free electrodynamics
in a flat corotating space-time following the approaches taken in
\cite{Carrasco:2020sxg,Most:2020ami}.\\
In the following, we will briefly summarize the equations of force-free electrodynamics,
the setup of the binary system, and the numerical implementation.

\subsection{Relativistic electrodynamics}
\label{sec:FFE}

In this section, we want to review the equations of general-relativistic
electrodynamics \cite{Baumgarte:2002vv,Palenzuela:2012my}.
The four-dimensional spacetime metric $g_{\mu\nu}$ can be expressed within
the 3+1 split as follows \cite{Gourgoulhon:2007ue},
\begin{align}
{\rm d}s^2=& g_{\mu\nu} {\rm d}x^\mu{\rm d}x^\nu \nonumber\\
          =& \left( -\alpha^2 + \beta_k \beta^k \right) {\rm d} t^2 
+ 2 \beta_k {\rm d} x^k {\rm d} t +\gamma_{ij} {\rm d}x^i {\rm d} x^j,
\end{align}
where $\alpha$ and $\beta^i$ are the lapse and shift respectively.
This decomposition introduces spatial hypersurfaces characterized by the
normal vector $n_\mu = \left(-\alpha,0,0,0\right) $.
Within this work, we will adopt a corotating flat frame, in which
$\alpha=1$ and $\gamma_{ij} = \delta_{ij}$, but non-zero shift $\beta^k$. We will specify the precise
choice of $\beta^k$ in Sec. \ref{sec:beta}, see also \cite{Carrasco:2020sxg,Most:2020ami}.
The dynamics of electromagnetic fields in relativity is governed by the
generalized form of Maxwell's equations
\cite{Palenzuela:2012my},
\begin{align}
  \nabla_\mu \left(F^{\mu\nu} + \phi g^{\mu\nu}\right) &= - 4 \pi
  \mathcal{J}^\nu   +\kappa_\phi n^\nu \phi, \label{eqn:Maxwell1_c} \\
  \nabla_\mu  \left(^\ast F^{\mu\nu} +\psi g^{\mu\nu}\right) &=
  +\kappa_\psi n^\nu \psi\,, \label{eqn:Maxwell2_c}\\
  \nabla_\mu \mathcal{J}^\mu &=0\,.
\end{align}
These equations are written in terms of the four-dimensional spacetime
metric $g_{\mu\nu}$, the field strength tensor $F^{\mu\nu}$ and its dual
$^\ast F^{\mu\nu} = \frac{1}{2} \varepsilon^{\mu\nu\kappa\lambda} F_{\kappa\lambda}$,
where $\varepsilon^{\mu\nu\kappa\lambda}$ is the Levi-Civita tensor.
In order to preserve the divergence-free condition of the magnetic field,
as well as consistency of the \rev{separately evolved} charge density
\rev{with} the electric field, we have
added divergence cleaning scalars $\phi$ and $\psi$ with corresponding
dissipation coefficients $\kappa_{\phi/\psi}$ \cite{Palenzuela:2008sf}.
We note that these extra fields vanish in the continuum limit, and that there presence
requires to separately impose charge conservation via Eq. \eqref{eqn:Maxwell2_c}.
For this choice of lapse, we obtain that the electromagnetic fields and
currents are given by \cite{Palenzuela:2012my}
\begin{align}
  B^\mu &= -n_\nu\,{^\ast} F^{\mu\nu}, \\
  E^\mu &= -n_\nu F^{\mu\nu}, \\
  q &= -n_\nu \mathcal{J}^\nu,\\
  J^\mu &= h^\mu_\nu\mathcal{J}^\nu,\\
\end{align}
where we have defined the electric field $E^\mu = \left(0, E^i\right)$, the magnetic field
$B^\mu = \left(0, B^i\right)$, the charge density $q$ and the current
$J^\mu = \left(0, J^i\right)$. We have further introduced the projector
$h_{\mu\nu}=g_{\mu\nu} + n_\mu n_\nu$.
Expanding out the Maxwell equations \eqref{eqn:Maxwell1_c} and
\eqref{eqn:Maxwell2_c} we obtain,
\begin{align}
  \partial_t  B^i + \partial_k \left(
  -\beta^k B^i + \varepsilon^{ikj} E_j  +  \phi \delta^{ik}\right)
  &= - B^k \partial_k \beta^i,\\
  \partial_t E^i + \partial_k \left(
  -\beta^k E^i - \varepsilon^{ikj} B_j  +  \psi \delta^{ik}\right)
  &= - E^k \partial_k \beta^i -4\pi J^i,\\
  \partial_t \psi + \partial_k \left( - \beta^k
    \psi +  E^k \right)
    &=   q - \kappa_\psi \psi.\label{eqn:Bevol}\\
  \partial_t \phi + \partial_k \left( - \beta^k
    \phi + B^k \right)
    &=  - \kappa_\phi \phi, \\
  \partial_t q + \partial_i \left(- \beta^i q + J^i
  \right) &=0. \label{eqn:qevol2}
\end{align}

\begin{table*}
  \centering
\begin{tabular}{|l|c|c|c|c|c|c|c|c|c|}
    \hline
    \hline
    &$f_{1}\left[\rm Hz\right]$  & $f_{2}\left[\rm Hz\right]$ &
    $\Omega/2\pi \left[\rm Hz\right]$ & $\mathcal{Q}_0\, \left[\rm km\right]$ &$B_{\rm 2}/B_{\rm 1}$ & 
     $\theta_1$ & $\theta_2$ & $\phi$ & $a\, \left[\rm km \right]$\\
    \hline
    \hline
	\texttt{A0} $\left(\downarrow\nearrow \, 0^\circ\right)$ & -- & -- & 259 & -- & 1.9& 0 & 180 & 0 & 52\\
	\texttt{A30} $\left(\downarrow\nearrow \, 30^\circ\right)$ & -- & --& 259 & -- & 1.9& 0 & 150 & 0 & 52\\
	\texttt{A60}$\left(\downarrow\nearrow \, 60^\circ\right)$ & -- & --& 259 & -- & 1.9& 0 & 120 & 0 & 52\\
	\texttt{A90}$\left(\downarrow\nearrow \, 90^\circ\right)$ & -- & --& 259 & -- & 1.9& 0 & 90 & 0 & 52\\
	\hline
	\texttt{O0} $\left(\nearrow\swarrow \right)$ & -- & --& 259& -- & 1.9& 0 & 120 & 0 & 52\\
	\texttt{O180} $\left(\nearrow\searrow \right)$& -- & --& 259 & -- & 1.9& 0 & 120 & 180 & 52\\
	\hline
        \texttt{Q}$\uparrow\uparrow$  & -- & --& 259 & 44 & 1& 0 & 30 & 0 & 52\\
	\texttt{Q}$^{\ast}\uparrow\uparrow$  & -- & --& 259 & 82 & 1& 0 & 30 & 0 & 52\\
	\texttt{Q}$\downarrow\downarrow$  & -- & --& 259 & 44 & 1& 0 & 120 & 0 & 52\\
	\hline
	\texttt{U0}$\left(\nwarrow\swarrow\, 1:1 \right)$  & --  & 100& 259 & -- & 1.9& 30 & 150 & 0 & 52\\
	\texttt{U3}$\left(\nwarrow\swarrow\, 1:3 \right)$ & --  & 100& 259 & -- & 0.636& 30 & 150 & 0 & 52\\
	\texttt{U9}$\left(\nwarrow\swarrow\, 1:9 \right)$ & --  & 100& 259 & -- & 0.121& 30 & 150 & 0 & 52\\
	\texttt{U27} $\left(\nwarrow\swarrow\, 1:27 \right)$ & -- & 100& 259 & -- & 0.071& 30& 150 & 0 & 52\\
	\hline
	\hline
  \end{tabular}
	\caption{Summary of the neutron star binaries
	  considered in this work. The columns denote the spin frequencies of
	  the stars $f_{\rm 1/2}$, the orbital spin frequency $\Omega$,
	  the relative maximum magnetic field strengths $B_{2}/B_1$ at the surface of
	  the stars, the inclination $\theta_{1/2}$, the relative
	  initial offset $\phi$ between the two stars and the separation
	  $a$. We further use $\mathcal{Q}_0$ to denote strength of the
	  quadrupolar \rev{component}, where applicable. In the decoupling limit between the electromagnetic field and
      the gravitational sector that we consider in this work, the magnetic
    field strength $B_1$ can be arbitrarily re-scaled, so that no explicit
  value needs to be provided. For the actual simulations, we use $B_1 = 1.9\,
 \times 10^{12}\, \rm G$.}
    \label{tab:initial}
\end{table*}
We recall that the relativistic energy momentum tensor for electric and
magnetic fields is given by \cite{Baumgarte:2002vv},
\begin{align}\label{eqn:TEM}
  4 \pi T_{\rm EM}^{\mu\nu} =& \frac{1}{2} \left[ E^2 + B^2 \right]g^{\mu\nu}
  + 2 n^{ (\mu} \varepsilon^{\nu) \kappa\lambda} E_\kappa B_\lambda
  \nonumber \\ 
  &-
  E^\mu E^\nu - B^\mu B^\nu\,,
\end{align}
where $\varepsilon^{\mu\nu\kappa\lambda}$ is the four-dimensional
Levi-Civita tensor and $2 A^{(\mu} B^{\nu)} = A^\mu B^\nu + B^\mu A^\nu$.
The electromagnetic energy density $\rho_{\rm EM}$ can then be recovered by
contracting $T_{\rm EM}^{\mu\nu}$ with the normal vector $n_\mu$,viz. ,
\begin{align}
  \rho_{\rm EM} = \frac{1}{8\pi} \left( B^2 + E^2 \right)\,.
  \label{eqn:rho_EM}
\end{align}
Similarly, the flux of electromagnetic energy {\it -- the Poynting flux -- }
can be expressed in the usual form,
\begin{align}
\left(S_{\rm EM}\right)_i = \frac{1}{4\pi} \varepsilon_{ijk} E^j B^k\,.
  \label{eqn:poynting}
\end{align}

It can be shown that the electromagnetic energy tensor obeys the following
conservation law \cite{Baumgarte:2002vv},
\begin{align}
  \nabla_\nu T^{\mu\nu}_{\rm EM} = F^{\mu\nu} J_\nu.
  \label{eqn:EM_T_conservation}
\end{align}
Together with the above definitions \eqref{eqn:rho_EM} and
\eqref{eqn:poynting}, we can re-express \eqref{eqn:EM_T_conservation}
as a conservation law for the electromagnetic energy density $\rho_{\rm EM}$,
\begin{align}
  \partial_t \rho_{\rm EM} + \partial_k \left( S^k_{\rm EM} - \beta^k
  \rho_{\rm EM} \right) = J_i E^i\,.
  \label{eqn:EM_T_31}
\end{align}
We can see that electromagnetic energy is conserved, except for dissipation as quantified by the rate $J_i E^i$.

\subsubsection{Force-free electrodynamics}

We model the dynamics of the highly conductive pair plasma outside of the neutron stars
using relativistic force-free electrodynamics.
This amounts to imposing the force-free conditions \cite{Komissarov:2004ms,Gruzinov:2004jc,Spitkovsky:2006np}
\begin{align}
    F^{\mu\nu} \mathcal{J}_\nu&=0\,,\\
    {^\ast}\!F^{\mu\nu} F_{\mu\nu} &=0\,,\\
    F^{\mu\nu} F_{\mu\nu} &>0\,.\\
\end{align}
We impose these conditions in the exterior of the neutron stars using the relaxation
approach of \cite{Alic:2012df}, i.e.
\begin{align}
  J^i = q \frac{\varepsilon^{ijk} E_j B_k}{B_l B^l} + \nu \left[\frac{E_j
  B^j}{B_l B^l}B^i + \frac{\chi \left( E_l E^l - B_l B^l \right)}{B_l B^l}
E^i \right],
  \label{eqn:J}
\end{align}
where \rev{$\nu^{-1}= 1.25\times 10^{-6} \Omega^{-1}$ is a short relaxation
time}, $\chi$ is
the Heaviside function, and $\Omega$ is the orbital angular speed, see Tab. \ref{tab:initial}.

This choice of force-free current allows for energy dissipation in the magnetosphere. Following Eq. \eqref{eqn:EM_T_31}, we can quantify this dissipation rate as
\begin{align}
  J_i E^i = \nu_1 \frac{\left( E_k B^k \right)^2}{B^2} + \nu_1 
  \chi \left( E^2 -B^2 \right) \frac{E^2}{B^2}.
  \label{eqn:JE_exp}
\end{align}
Hence, dissipation in our simulations scales with the energy density of the
parallel electric field, $E_\parallel=\left(E_i B^i\right)/\sqrt{B^2}$.

\subsection{Quasi-circular inspiral}
\label{sec:beta}


Since we want to simulate the interaction of the magnetospheres of 
two neutron stars in a close contact binary, we also need to model the stars as well as
their orbital and rotational motion. 
For simplicity, we adopt spherical conductors with a radius of $14.76\,
\rm km$ on a circular orbit. Our results are insensitive to small variations of the radius
within currently favored uncertainties, see e.g. \cite{Raaijmakers:2019dks}. 
In this quasi-adiabatic inspiral, we neglect
changes in the orbital separation, $a$, which is a good approximation at moderate
distances, since the separation changes only slowly compared to the orbital
period \cite{Peters:1964zz}.
Following Refs. \cite{Carrasco:2020sxg} and \cite{Most:2020ami}, we model the
rotation of the sphere in terms of a local spin frequency $\omega_{1/2}$ of
the spheres around \rev{their} axis, and a global orbital motion in terms of the
orbital frequency $\Omega$. We can then split the advected fluid velocity
$u^i/u^0 = v^i - \beta^i$ into a global shift and three-dimensional
velocity part $v^i$. For these, we adopt 
\begin{align}
  \beta^i =&\, - \varepsilon^{ijk} \Omega \hat{z}_j x_k, \\
  v^i =&\,  \varepsilon^{ijk} \omega \hat{z}_j \left( x_k - x^{ 1/2}_k
  \right) - \varepsilon^{ijk} \Omega \hat{z}_j
  x^{1/2}_k, \label{eqn:vinside}\\
  u^i/u^0 =&\, v^i - \beta^i \nonumber\\
  =&\, \varepsilon^{ijk} \left( \Omega + \omega \right) \hat{z}_j
  \left( x_k - x^{ 1/2}_k \right) ,
  \label{eqn:uiu0}
\end{align}
where we have denoted the coordinate $x_k$ and location of the stellar
centers $x^{1/2}_k$, where $\hat{z}_j$ is the unit vector perpendicular to the orbital plane. 
\rev{This construction results in a flat metric, describing a corotating
frame uniformly spinning at the orbital angular frequency $\Omega$.}
We can now see that the bulk motion of the stars is
purely rotational around their spin axis, which we assume to be aligned
with the orbital axis. The relative rotation rate is given by the sum
$\Omega + \omega$, where $\Omega = -\omega$ would correspond to a tidally
locked, corotational system, and $\Omega = -2 \omega$ to an irrotational
binary. Since we evolve the electric and magnetic fields also inside the
conducting spheres, we find that such a prescription leads to a major
reduction in spurious behavior (e.g. artificial dissipation) caused by the motion of the stars.
Physically, the inside of a neutron star will be (almost) infinitely
conductive, so that the electric field is well approximated by the
limit of ideal magnetohydrodynamics 
\begin{align}
    E^i = -\varepsilon^{ijk} v_j B_k\,,
    \label{eqn:idealMHD}
\end{align}
with $v_j$ given by 
Eq. \eqref{eqn:vinside} inside the conducting spheres.

\subsection{Magnetic field geometry}

We model the geometry of the magnetic fields 
of the neutron stars either as two dipoles or as
quadrudipoles \cite{Gralla:2017nbw}.
The latter are loosely motivated by recent observations of PSR J0030+0451
by the NICER collaboration \cite{Bilous:2019knh}.
More precisely, we use the following expressions for the  vector potentials
$A_\mu = \left(0, A_i\right)$, where 
\begin{align}
  A_i =  \left(-\frac{y - y_{1/2}}{\varpi^2} A_\phi , \frac{x - x_{1/2}}{\varpi^2}
    A_\phi,0\right)
\end{align}
is determined in cylindrical
coordinates $\left(\varpi\,,\phi\,,z\right)$ centered on each star.
For the dipole, we use \cite{Shibata:2011fj},
\begin{align}
  A_\phi = A_0 \frac{\varpi_0 \varpi^2}{\left( \varpi^2 + z^2 + \delta^2
    \right)^{3/2}}\,,
  \label{eqn:A_dipole}
\end{align}
whereas we use the following for the quadrudipole \cite{Gralla:2017nbw},
\begin{align}
  A_\phi = A_0 \frac{\varpi_0 \varpi^2}{\left( \varpi^2 + z^2 + \delta^2 
    \right)^{3/2}} \left(1 - \mathcal{Q}_0
    \frac{z}{\varpi^2 + z^2 + \delta^2}\right)\,,
  \label{eqn:A_quad}
\end{align}
where $\delta$ is a small number to regularize the divergence of the vector
potential at $\varpi=z=0$.
Additionally, we have introduce the characteristic quadrupolar length scale
$\mathcal{Q}_0$, which will determine the physical extent of the quadrupole
field.
We then compute the magnetic field using the curl-expression for the
magnetic field, i.e. $B^i = \varepsilon^{ijk}\partial_j A_k$. Different
inclinations of the dipoles can simply be incorporated by rotating the
resulting vector field. We point out that the use of a divergence cleaning
approach simplifies the discrete computation of the initial magnetic field,
and we opt to evaluate the magnetic field expression analytically.
Furthermore, since we are evolving the interior of the stars, the 
expression above are only used at the initial time. The remainder of the
simulation proceeds self-consistently from there onward.
A summary of the initial conditions used in this work is provided in 
Tab. \ref{tab:initial}.

\subsection{Numerical implementation}

We solve Maxwell's equations \eqref{eqn:Maxwell1_c} and
\eqref{eqn:Maxwell2_c} numerically following standard
approaches used in the numerical relativity community to solve the 
force-free electrodynamics system in general space-times \cite{Alic:2012df,Palenzuela:2012my}.
All spatial gradients are evaluated using a
fourth-order accurate finite volume scheme \cite{mccorquodale2011high}
based on WENO-Z reconstruction \cite{Borges2008}. The flux terms are
computed using a simple Rusanov Riemann solver \cite{Rusanov1961a}, with
the fluxes being computed for each spatial direction separately.
Instead of solving the Riemann problem in the global frame, we transform to
a local aligned tetrad frame \cite{White:2015omx,Kiuchi:2022ubj} in order to best address the issues raised by
diverging characteristics $c_{\pm}$.

Overall this scheme is similar to other schemes used for binary neutron
star merger simulations in full numerical relativity \cite{Most:2019kfe}.
Such high-order schemes have been shown to be beneficial for the study of
force-free electrodynamics \cite{Mahlmann:2020yxn}.
We point out the use of the effective dissipative current given in Eq. \eqref{eqn:J} does allow for reconnection to occur at resistivities
above the grid scale \cite{Ripperda:2021pzt,Mahlmann:2021yws}, with the current sheet
thickness set by the $\nu_1$ parameter. 

Due to the stiffness of the relaxation current, we need to adopt an
implicit time evolution scheme to ensure stable evolutions. Following
\cite{Alic:2012df,Palenzuela:2012my}, we adopt the third-order accurate
implicit-explicit (IMEX) strong stability preserving RK-SSP(4,4,3) scheme
\cite{pareschi_2005_ier}.  

The equations are solved on a discrete computational grid consisting 
of a nested box-in-box refinement structure. The grid extends to $\simeq
1200\, \rm km$ in each direction and the innermost grid extends to $\simeq
50\, \rm km$ per direction.  The computational grid is provided by a set of
nested boxes using the AMReX \cite{amrex} highly parallel adaptive
mesh-refinement framework, on which \texttt{GReX} is built. In total,
we use $8$ refinement levels with a highest resolution of $260 \,{\rm m}$.

We enforce \eqref{eqn:idealMHD} by imposing an
ad-hoc current
\begin{align}
  J^i_{\rm star} = K (E^i + \varepsilon^{ijk} v_j B_k),
  \label{eqn:Jstar}
\end{align}
where $K$ is fixed based on the Runge-Kutta time-stepping such that $J^i_{\rm star}$
results in the exact enforcement of the ideal MHD condition at each substep
of the numerical time integrator.\\

Since the angular component of the global shift scales as $\beta^\phi = -
\Omega \varpi$, where $\varpi$ is the cylindrical orbital radius, it is
apparent that outside of the light cylinder of the orbit $\beta^\phi >1$,
hence causing apparent superluminal motion on the grid. 
More formally, the use of the corotating frame imposes
characteristic speeds of the system
\begin{align}
  c_\pm = 1 \pm \beta^i\,,
  \label{eqn:cpm}
\end{align}
where $\beta^i$ is the component of the shift in the $i$-th direction.
While apparently superluminal characteristics do not violate causality, 
they impose a severe restriction on the numerical time stepping algorithm, since the largest
(apparent) velocity on the grid would no longer be the speed of light
$c=1$, but the corotation speed at the boundary of the grid, $\Omega
D/2$, where $D$ is the box size of the domain.\\

\begin{figure*}
  \centering
  \includegraphics[width=0.95\textwidth]{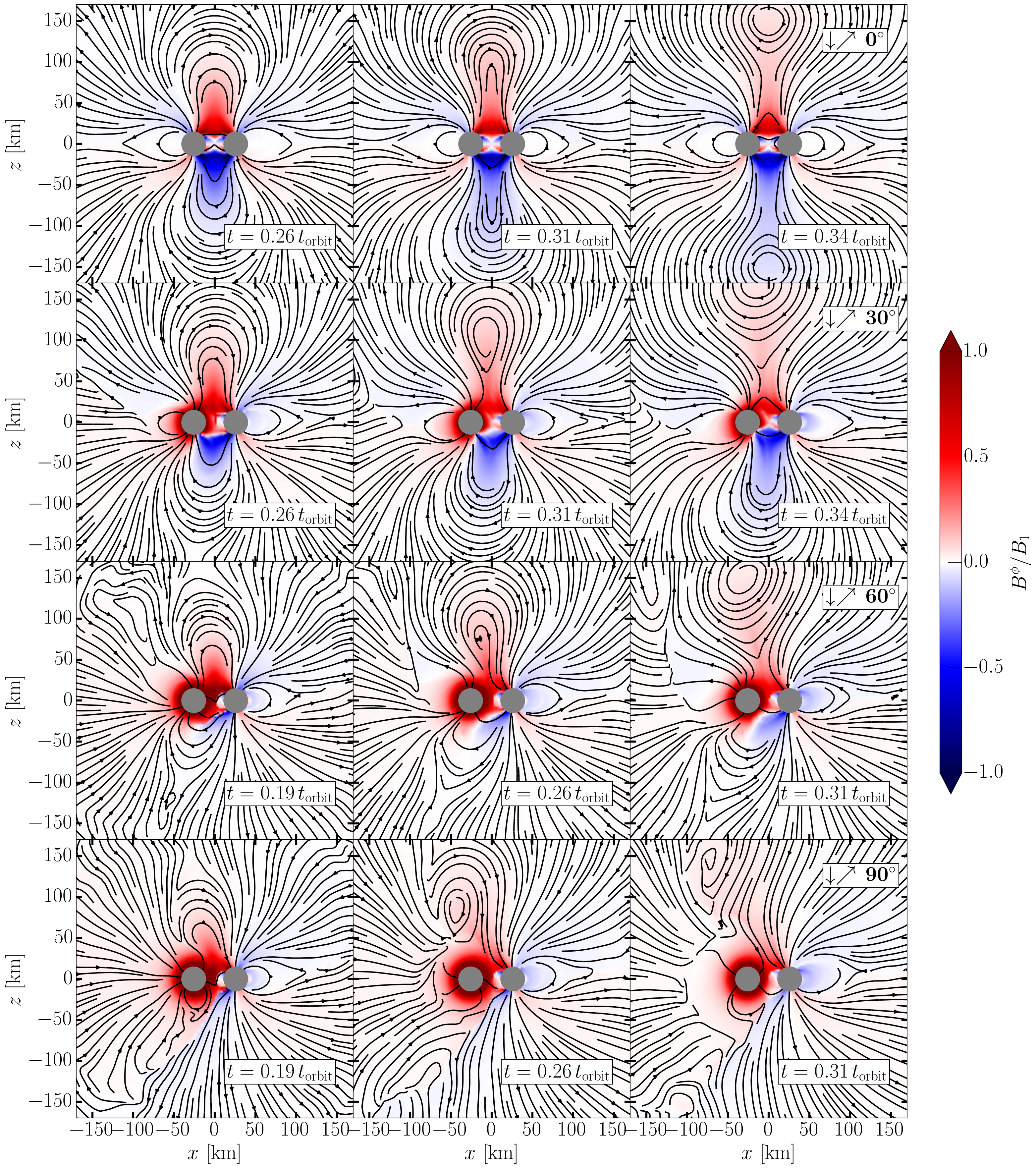}
  \caption{Pre-merger flaring process in the late inspiral of a coalescing binary neutron star
  system. The relative rotation of the two neutron stars (gray), leads to a
twist in the common magnetosphere indicated by the out-of-plane magnetic
field component $B^\phi$. The magnetic field lines correspond to a
projection of the field onto the meridional plane. 
Times are stated in units of the orbital time scale $t_{\rm orbit}$. The rows correspond to different inclinations of the
magnetic moment of the primary star (models \texttt{A0}--\texttt{A90}), relative to the a reference value $B_1$.}
  \label{fig:alignment}
\end{figure*}

\section{Results}
\label{sec:results}

\begin{figure*}
  \centering
  \includegraphics[width=0.45\textwidth]{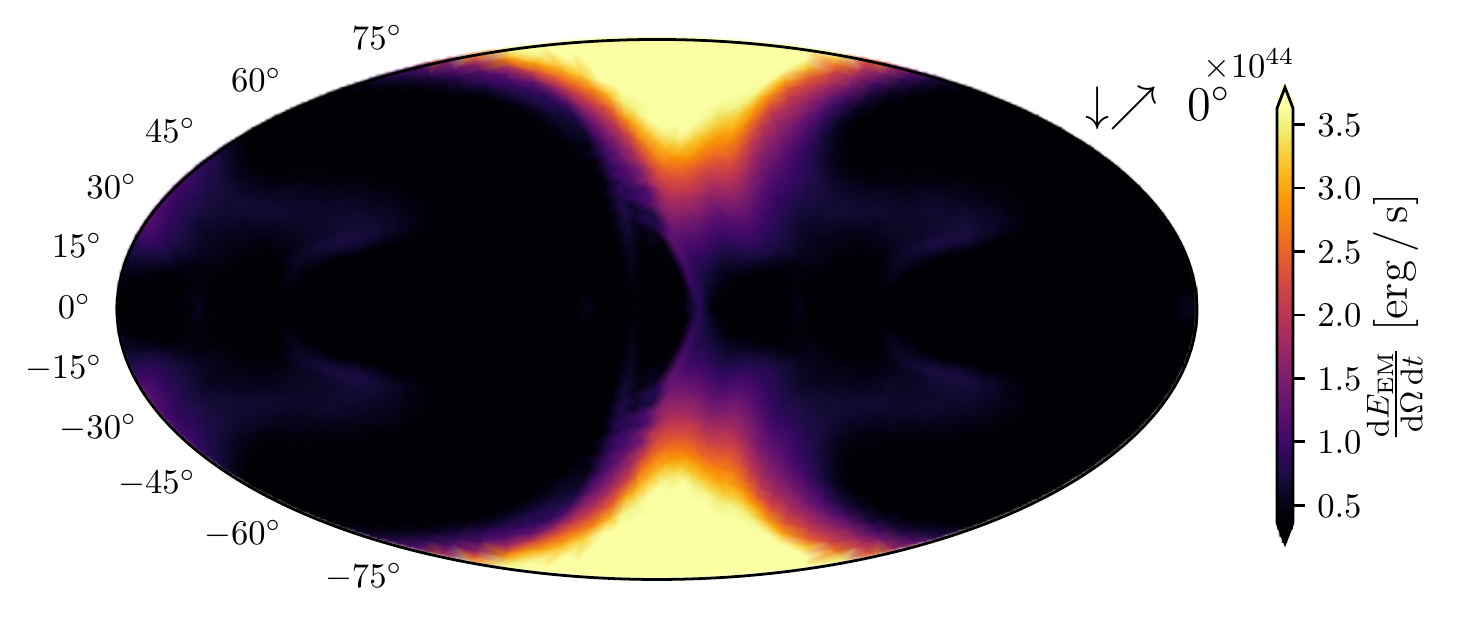}
  \includegraphics[width=0.45\textwidth]{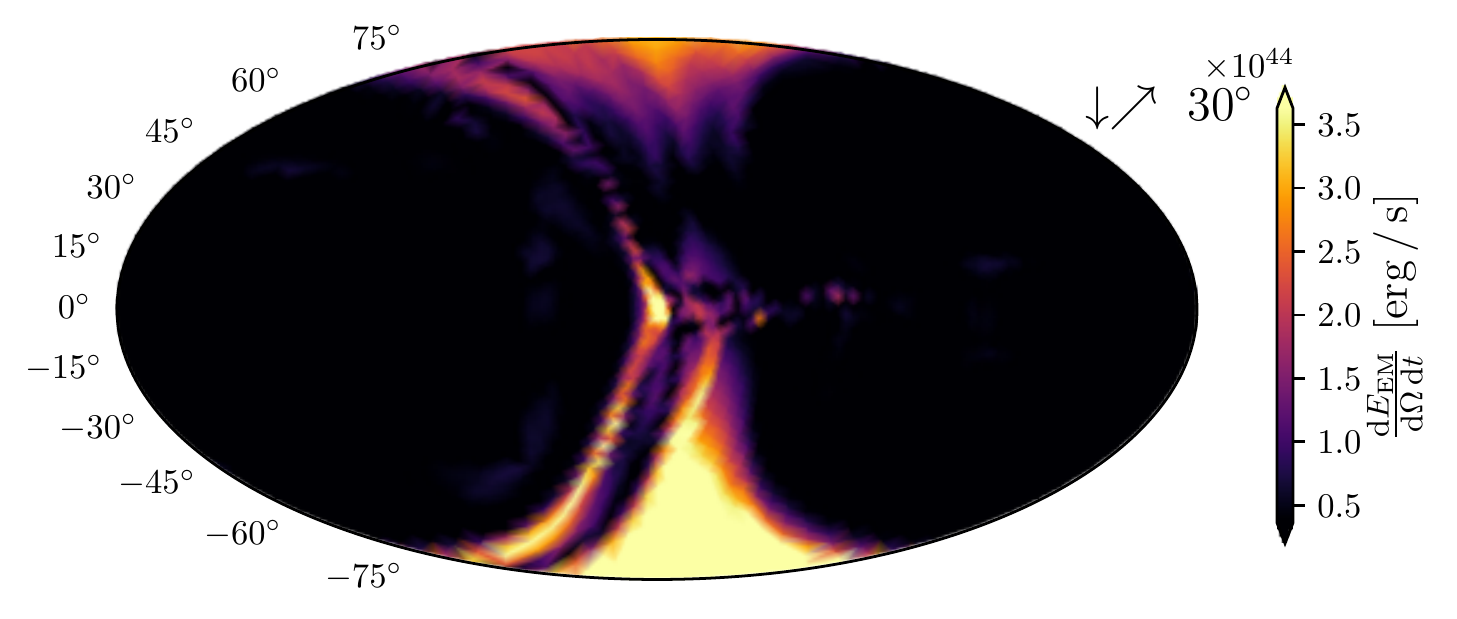}\\
  \includegraphics[width=0.45\textwidth]{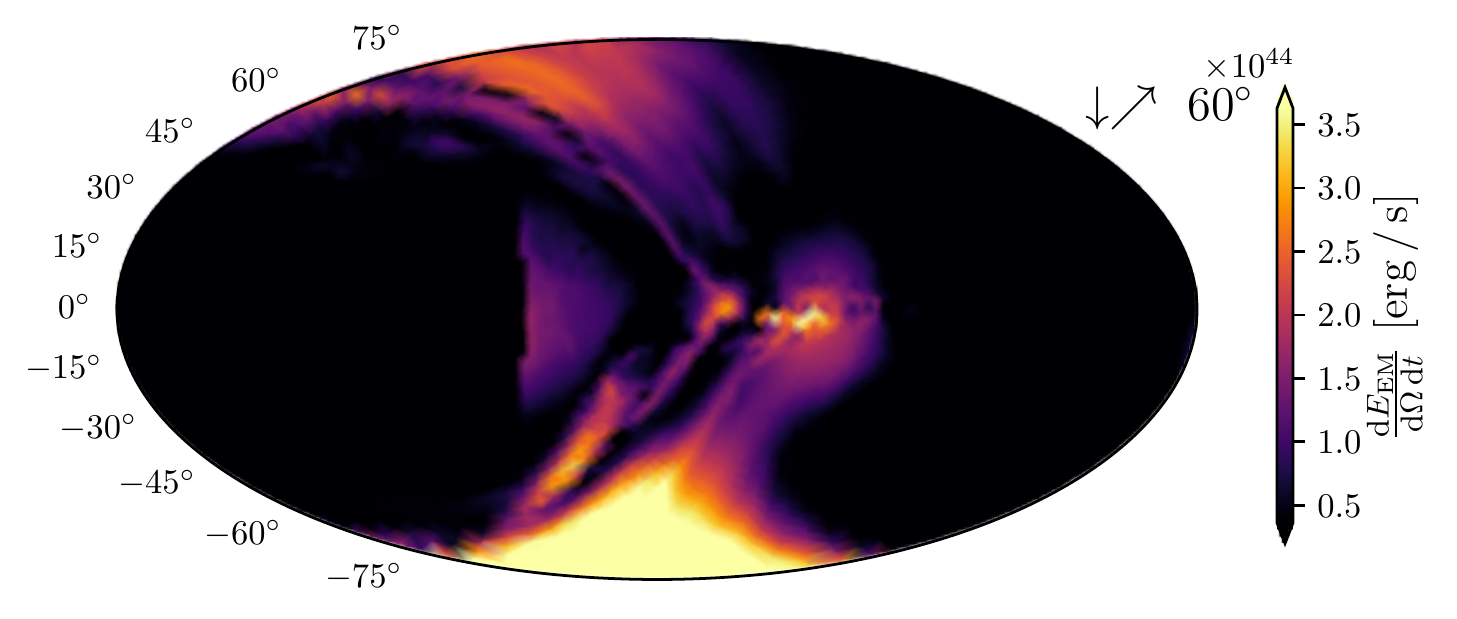}
  \includegraphics[width=0.45\textwidth]{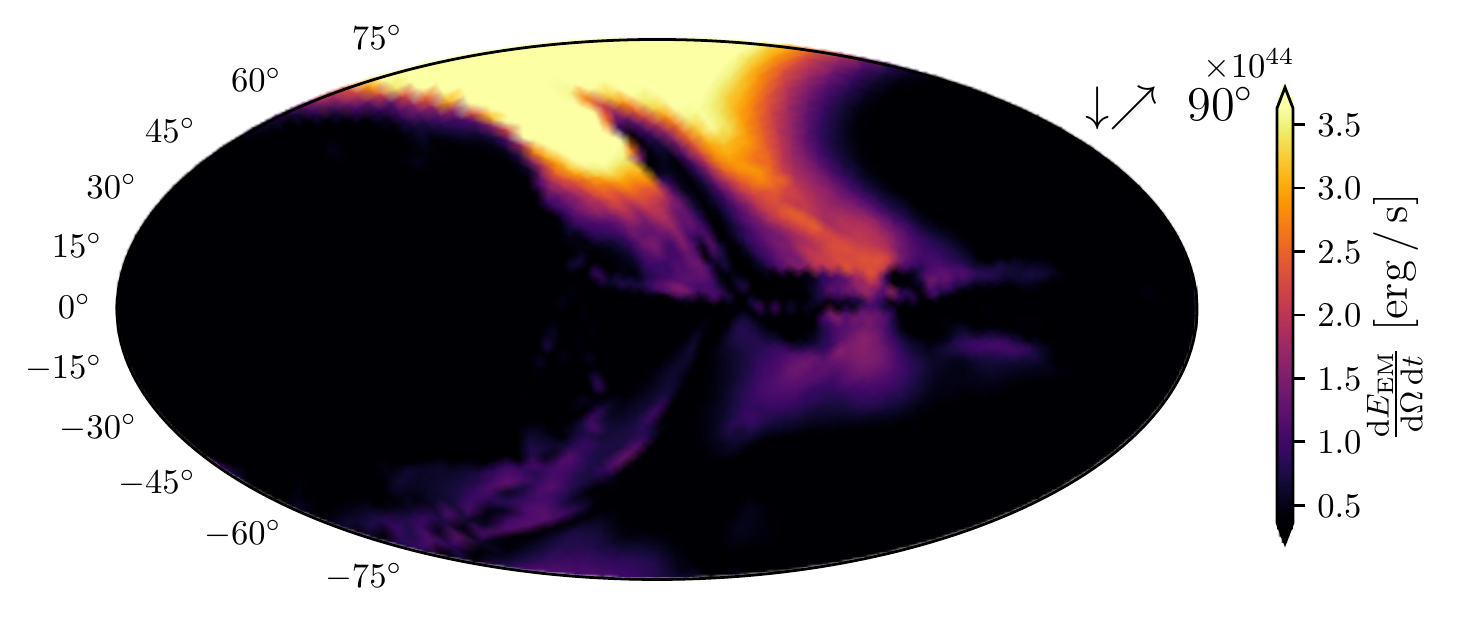}
  \caption{Time-averaged radial Poynting flux ${{\rm d}E_{\rm
  EM}}/{\left({\rm d} t {\rm d} \Omega \right)}$ projected onto a sphere at
  $r=370\, \rm km$ from the origin. The time-averaging was done over one
flaring period, and includes the removal of the energy flux of the quiescent state of the system.
Shown are the models of equal magnetic field strengths, but varying dipole inclinations (\texttt{A0}--\texttt{A90}).
The plots adopt a Mollweide projection and are aligned with the orbital rotation axis.}
  \label{fig:poynting2D_alignment}
\end{figure*}

In the following, we will present the results of our numerical
investigation of pre-merger flaring in a coalescing binary neutron star system.
In particular, we will demonstrate the feasibility of the flaring
mechanism for a variety of magnetic field topologies and orbital parameters.
In doing so, we will proceed in two steps. First, in
Sec.\ref{sec:alignment}-\ref{sec:phase} we will
demonstrate the global dynamics of the flaring process for various binary
configurations. We will also show how the emission geometry correlates with
the magnetic field topology.
In the second part, Sec. \ref{sec:luminosity}, we will estimate the amount of
electromagnetic energy emitted and dissipated in this process. This will help 
to make quantitative predictions about the potential detectability of electromagnetic
transients associated with the flaring process.\\

In all of this work, we consider a series of neutron star binaries at fixed orbital separation
$a=52\, \rm$ km, \rev{which} corresponds to the late inspiral of the binary,
i.e. $\lesssim 10$ orbits before merger. This close interaction was
identified as the most promising, since the luminosity of the flaring
process was shown to scale as $a^{-7/2}$ with the orbital separation $a$
\cite{Most:2020ami}.
We further fix\footnote{In the absence
of back reaction of the gravitational sector (the simulations are performed
in flat spacetime) on the electromagnetic fields, the Maxwell equations
\eqref{eqn:Maxwell1_c} and \eqref{eqn:Maxwell2_c} with the current
\eqref{eqn:J} become invariant under a global constant re-scaling of the magnetic and electric
field. This allows for a simple a posteriori rescaling of the reference
field strength.} the magnetic field strength at the pole of the primary neutron star
to $B_1 = 1.9\times 10^{12}\, \rm G$, see Tab. \ref{tab:initial}. 

\subsection{Effects of dipole inclination}
\label{sec:alignment}
\begin{figure*}
  \centering
  \includegraphics[width=0.96\textwidth]{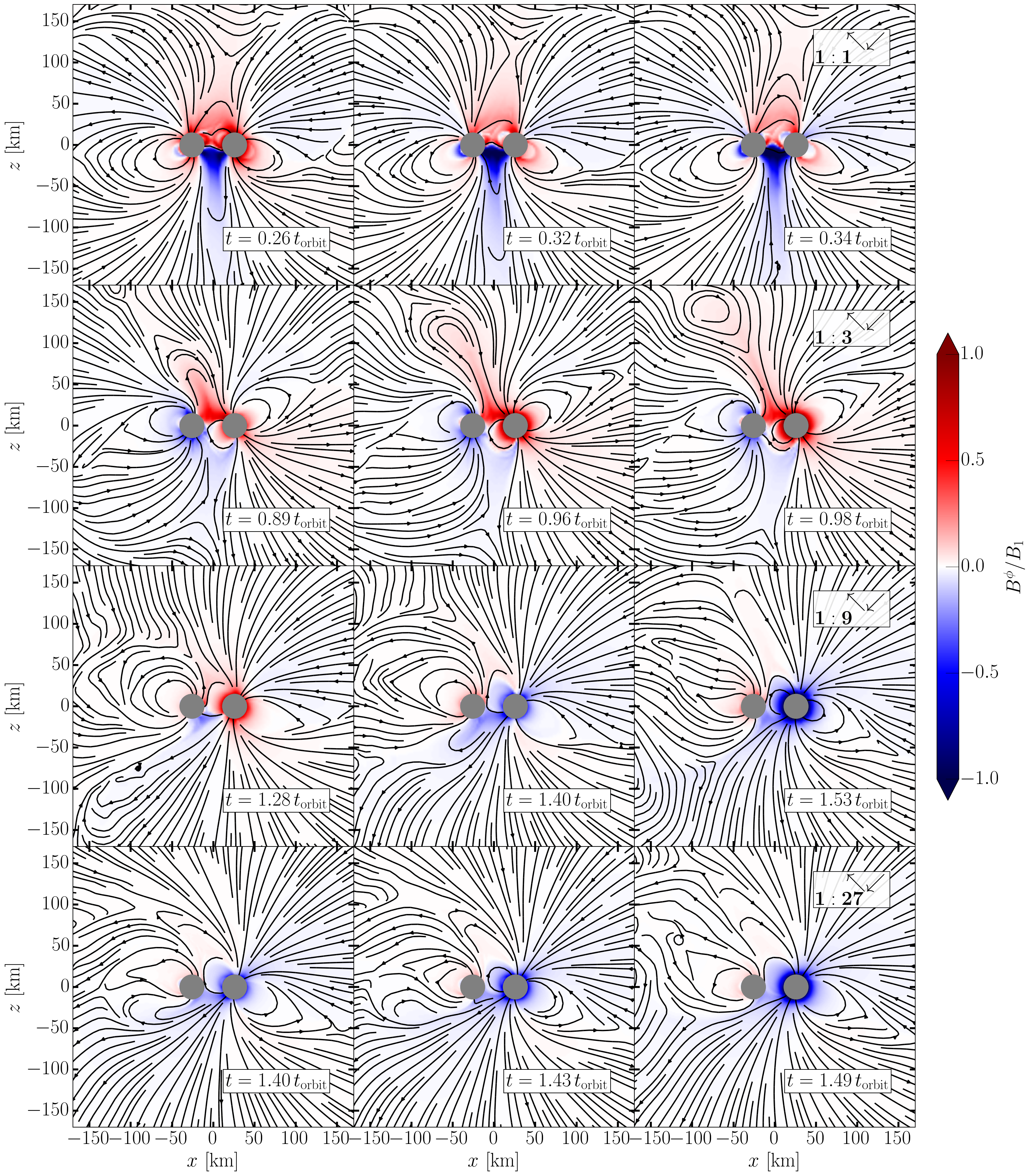}
  \caption{Global flaring dynamics for asymmetric field strengths in a
    close contact neutron star binary with misaligned dipole magnetic fields.
    Same as Fig. \ref{fig:alignment}, but for the unequal
  field strength models (\texttt{U0}--\texttt{U27}). }
  \label{fig:unequal}
\end{figure*}

\begin{figure*}
  \centering
  \includegraphics[width=0.45\textwidth]{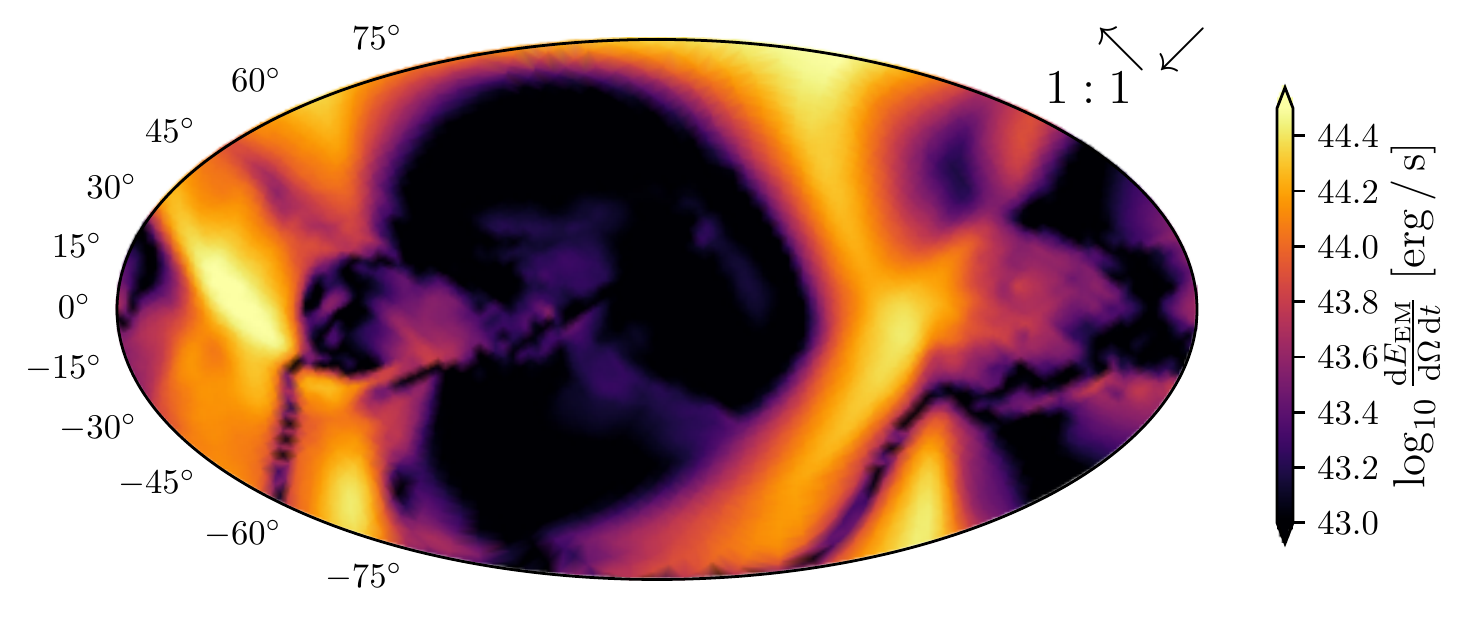}
  \includegraphics[width=0.45\textwidth]{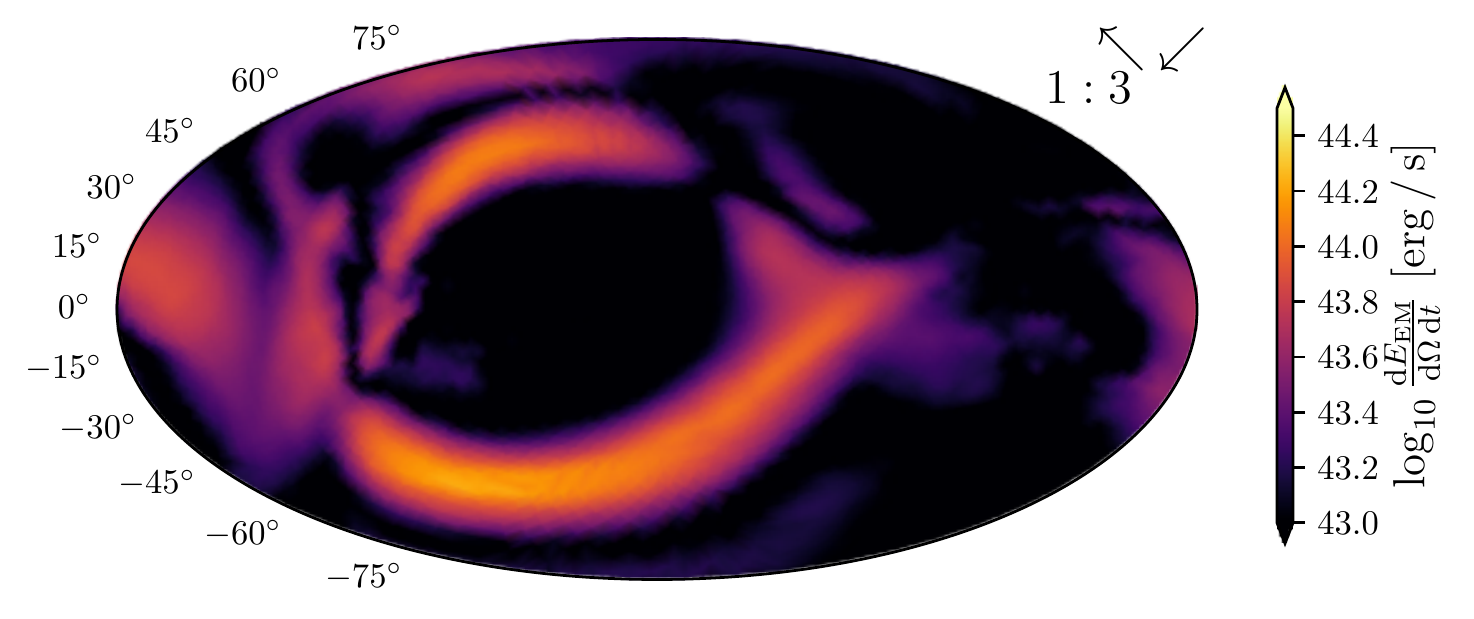}\\
  \includegraphics[width=0.45\textwidth]{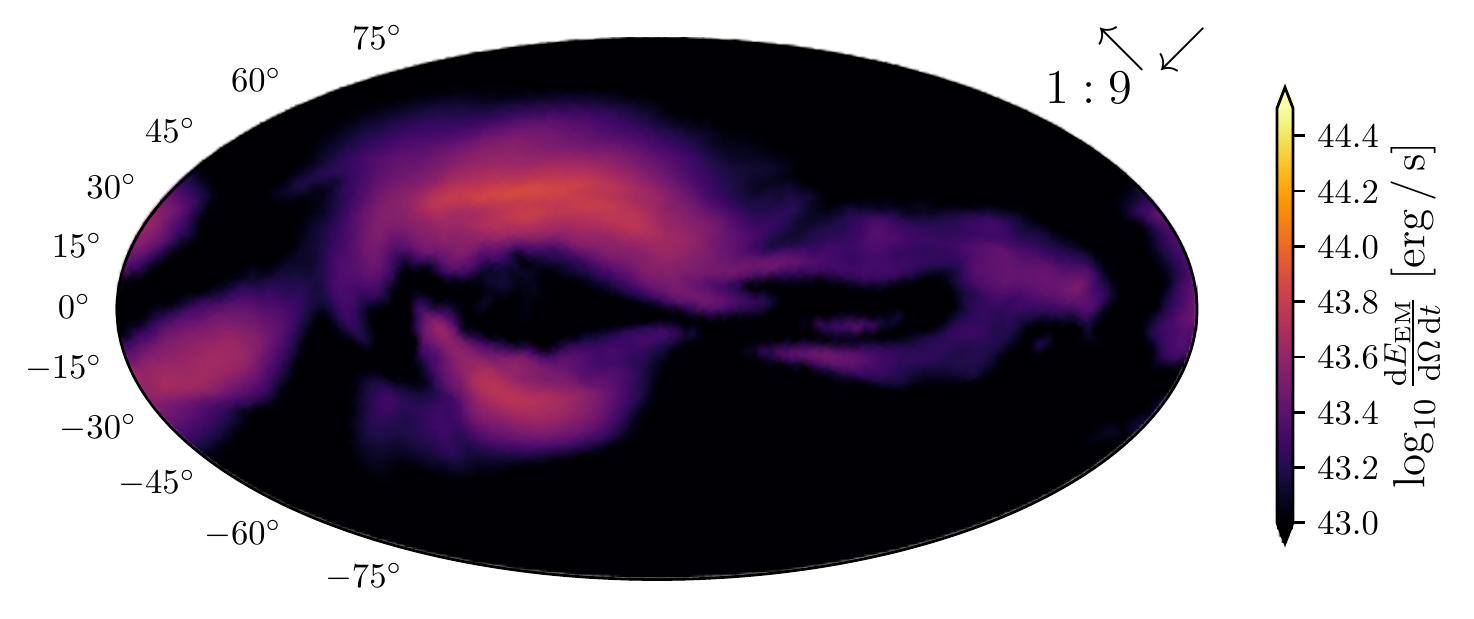}
  \includegraphics[width=0.45\textwidth]{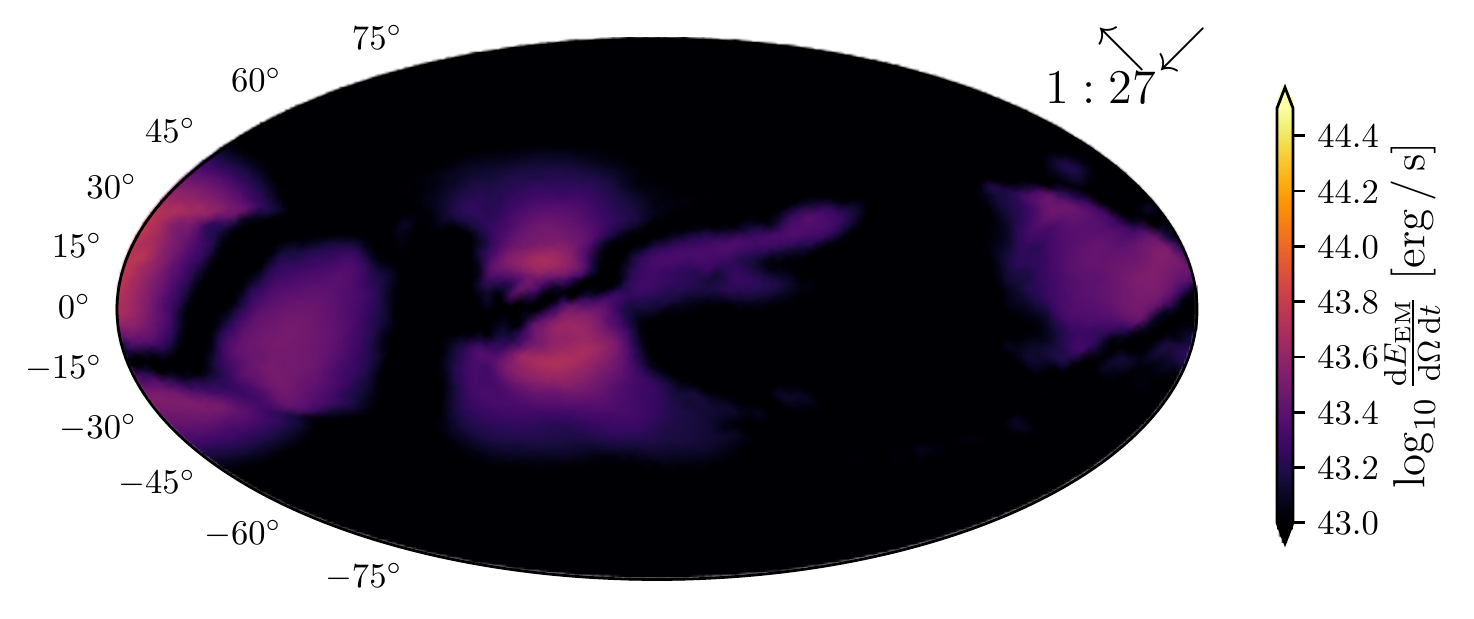}
  \\
  \caption{Time-averaged radial Poynting flux $\frac{{\rm d}E_{\rm
  EM}}{{\rm d} t\,{\rm d} \Omega}$ projected onto a sphere at
  $r=370\, \rm km$ from the origin. Same as Fig.
  \ref{fig:poynting2D_alignment}, but for different field strengths on the
  secondary star (see models \texttt{U0}--\texttt{U27} in Tab.
  \ref{tab:initial}).}
  \label{fig:poynting2D_uneq}
\end{figure*}

In order to better understand the flaring dynamics, we start out by
reviewing the flaring mechanism first described in \cite{Most:2020ami}.
For this we consider two equally magnetized neutron stars with magnetic
moments pointing in opposite directions (model \texttt{A0}), see also the
first row of Fig. \ref{fig:alignment}. Since the orbital motion will have
reignited the pair-cascade long before the final orbits of the inspiral, 
the ambient medium is filled with a force-free pair-plasma
\cite{Hansen:2000am,Lyutikov:2018nti,Wada:2020kha}.
As a consequence, the oppositely pointing magnetic field lines can
reconnect, and form closed loops (top left panel of Fig.
\ref{fig:alignment}). Since all binaries considered here are irrotational and not tidally
locked \cite{Bildsten:1992my},
the stars rotate relative to each other when seen in the orbital corotating
frame, see also Ref. \cite{Cherkis:2021vto}.
The relative rotation of the two stars builds up a twist in the connecting
magnetic flux tube, transferring (orbital) rotational energy into the
magnetosphere. This twist is mediated by launching Alfven waves along
the flux tube \cite{Parfrey:2013gza}, which strictly requires the orbital
period to be \rev{larger}
than the Alfven wave travel time, hence,
demanding that the twisting happens inside the orbital light cylinder
with radius $r_L= c/\Omega = 184\, {\rm{km}} \gg a$ for all models.
Once magnetic field lines become more and more twisted, \rev{a solar flare
like eruption will occur, which features a trailing} current sheet that
will form between them (top middle panel Fig. \ref{fig:alignment}).
As the current sheet become tearing-mode unstable, 
reconnection will lead to the formation of plasmoids (not shown in Fig. \ref{fig:alignment}) and 
the disruption of the sheet \cite{Most:2020ami}.
This will in turn lead to the detachment of a magnetic bubble
 (top right panel of Fig. \ref{fig:alignment}) \rev{being able to shock the
 ambient medium further away from the system }
and potentially lead to the emission of radio
emission via a synchrotron maser process \cite{Beloborodov:2017juh,Metzger:2019una}. 
Additionally, the dissipation in the current sheet might lead to \rev{X-ray
emission} \cite{Beloborodov:2020ylo}.\\
The amount of available magnetic flux that can be twisted and reconnected
will depend on the relative inclination of the magnetic fields. To clarify
this dependence, we have performed a set of simulations (models
\texttt{A30},  \texttt{A60} and  \texttt{A90}) with varying relative
inclination, where we have defined $0^\circ$ inclination as the case of
oppositely aligned magnetic moments. To avoid the issue of relative alignment of
the two dipoles, see Sec. \ref{sec:alignment}, we focus on the scenario where only one of the dipoles
is inclined.
These cases are shown in the lower three rows of Fig. \ref{fig:alignment}.
In all cases, we can see that a sufficient number of magnetic flux tubes is available
for twisting. However, the effective twists become weaker for increasing
inclinations. For model  \texttt{A90}, the twisted flux tube is not in plane
and the emission is not symmetric (i.e. along the rotation axis), which
\rev{ differs from model  \texttt{A0}, for which} no
inclination and, thus, symmetry breaking is present. Crucially, the flaring
mechanism \rev{operates} in either case. That is, once an over twisting sets in,
reconnection at the base of the loop snaps the field lines, and a detached
magnetic bubble is launched from the merger site. This happens even in the
case of strong misalignment, where both bubbles and trailing current sheets
are present in the northern and southern hemisphere (e.g., bottom right
panel of Fig. \ref{fig:alignment}).\\
\begin{figure*}[t]
  \centering
  \includegraphics[width=0.95\textwidth]{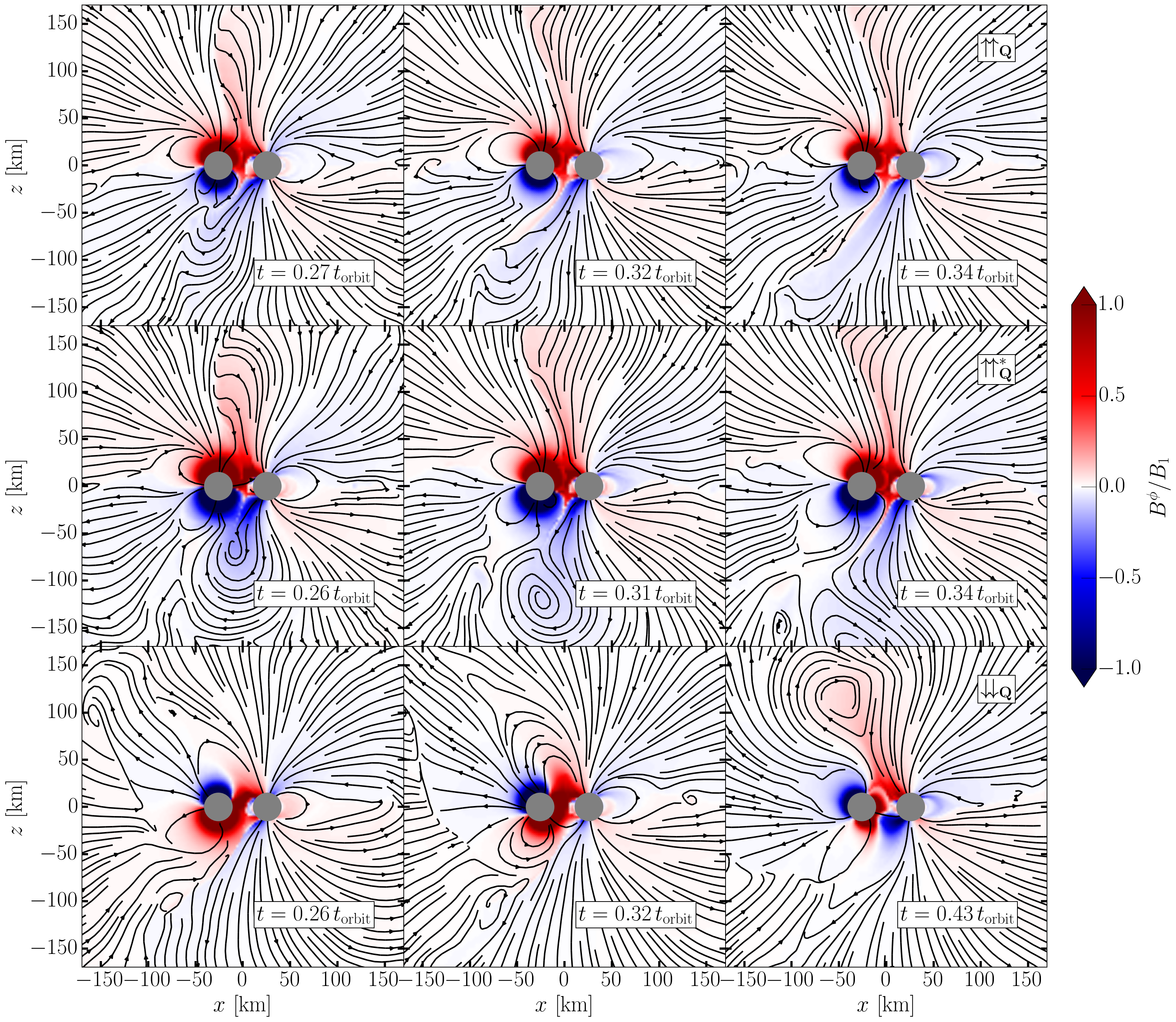}
  \caption{Global flaring dynamics for quadrudipolar magnetic field
    configurations. Same as Fig. \ref{fig:alignment} but for models, 
    \texttt{Q}$\uparrow\uparrow$, \texttt{Q}$^{\ast}\uparrow\uparrow$ and
    \texttt{Q}$\downarrow\downarrow$.
  }
  \label{fig:quadrudipole}
\end{figure*}

In order to better quantify the emission geometry of the flares, we study
the projection of a single flaring event onto a sphere corotating with the
orbital angular velocity $\Omega$. Since we are simulating within the
orbital corotating frame, the sphere remains fixed w.r.t. the stars.
The sphere is located about $370\,{\rm km} \simeq 7.1 a$ from the center-of-mass of
the binary. Following our discussion in Sec. \ref{sec:analysis}, we
quantify the flare in terms of its Poynting flux $S^i_{\rm EM}$, see
\eqref{eqn:poynting}. Instead of showing the spherical projection of the
Poynting flux at a fixed time, we isolate a single flare, and then
integrate the radial Poynting flux $S^r=S^\mu_{\rm EM} r^\nu g_{\mu\nu}/r$
in time over one flaring episode, where $r^\nu$ is spherical radius vector
with length $r$. Furthermore, we subtract the
time-averaged Poynting flux before and after the flaring event.
This way, we can approximately remove the intrinsic \rev{Poynting} flux due to rotation
of the stars in the corotating frame.
The resulting distributions are shown in Fig.\ref{fig:poynting2D_alignment}.
Starting from the fully anti-aligned system (top left), we can see that in
this geometry the twisting of the connected flux tube happens roughly in
the meridional plane. After the end points of the loop have reconnected, a
flare is then launched, propagating entirely in the polar direction, and
hence crossing the sphere exactly at the poles. This is the same
situation investigated previously in Ref. \cite{Most:2020ami}.
With varying inclination, the geometry of the flares becomes more
\rev{non}-axisymmetric. While in the perfectly anti-aligned cases, the same flux is
available in both hemispheres and can be equally twisted, this no longer
holds for the inclined models. 
In fact, one can clearly see that the flares
emerge from primarily reconnecting magnetic flux in the upper of the
lower hemisphere (see e.g., bottom row of Fig. \ref{fig:poynting2D_alignment}). For the $90^\circ$ model,
$\texttt{A90}$, the flaring happens exclusively in either hemisphere.
Since the flaring is periodic with half the orbital period in this case, the
system will be flaring in each hemispheric direction with a full orbital
period.
Additionally, the flares become more and more axisymmetric, being beamed
into the direction of the inclined secondary (left), with the flux going as
low as $\theta=60^\circ$ latitude, compared to only $\theta=75^\circ$ latitude for the
anti-aligned $(0^\circ)$ model. We also point out that the residual
Poynting flux on the equatorial plane, $\theta =0^\circ$, is a result of
imperfect removal of the constant Poynting flux from the rotation of the
inclined magnetic field. Therefore, this part is not associated with flaring, \rev{and instead} does also not overlap with the flare, which propagates
predominantly in the polar direction. 
While on first glance the energy flux of the flare appears comparable in all
cases, we will quantify the differences in Sec. \ref{sec:luminosity}.

\subsection{Effects of unequal magnetization}
\label{sec:unequal}

Following up on our previous discussion of magnetic field inclination
effects in Sec. \ref{sec:alignment}, we now focus on the general case
were the magnetic fields are misaligned and not equal in field strength. 
We focus on a close binary having two dipolar
magnetic moments misaligned by $30^\circ$ with their respective stellar
spin axis. Different from the previous cases (\texttt{A0}--\texttt{A90}),
the secondary has a weaker field strength $B_2 = B_1 / 3^n$ for $n=\left[
0,1,2,3 \right]$ and a moderate spin of $f_2 = 100\, \rm Hz$.
We follow the same logic as in Sec.  \ref{sec:alignment} and begin by first
discussing the global flaring properties shown in Fig. \ref{fig:unequal},
before moving on to describing the emission geometry of the Poynting fluxes in Fig.
\ref{fig:poynting2D_uneq}.\\
\rev{We now describe} the overall flaring dynamics shown in Fig.
\ref{fig:unequal}. Starting with the case of equally strong dipolar fields
(top row) we can see that the flaring process proceeds in the same way as
for the perfectly anti-aligned fields (top row, Fig. \ref{fig:alignment}).
A twist will build up in the common magnetospheres, leading the connected
flux tubes to reconnect and flare. Because of the different inclination of
both stars the flaring will largely proceed in the polar direction but out
of the meridional plane. Once we start to decrease the magnetic field
strength of the secondary by a factor three (second row), we can see that
the flaring happens in the same way (this time even being more in-plane).
However, because of the weaker field on the secondary, the twist will build
up closer to its surface, while the emerging loop of field lines bends closer around the secondary
star (right column, second row).
As the flaring proceeds, we can see that ejected flare is inclined towards the more weakly magnetized
secondary (right column).\\ When decreasing the field strength by a factor nine (third row)
this trend easily continues. In fact, the twisted loop is
now much closer to the secondary (middle panel), and the twist -- expressed
in terms of the out-of-plane field $B^\phi$ -- becomes weaker. The flares
are heavily bent towards the secondary. Finally, if we decrease the field
strength further (bottom row), flares are still being emitted. However, the
twist builds up almost at the surface of the secondary, with the twisted
flux tubes protruding far behind the secondary (left panel).
\begin{figure*}
  \centering
  \includegraphics[width=0.45\textwidth]{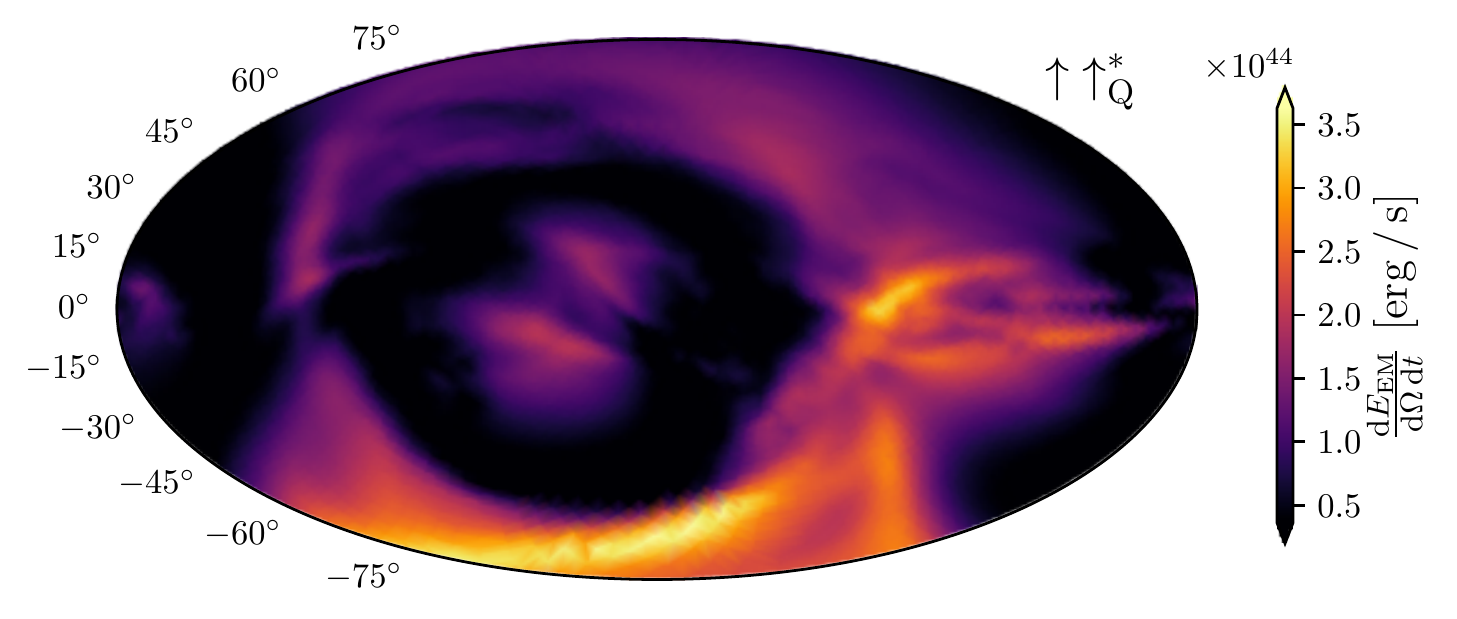}
  \includegraphics[width=0.45\textwidth]{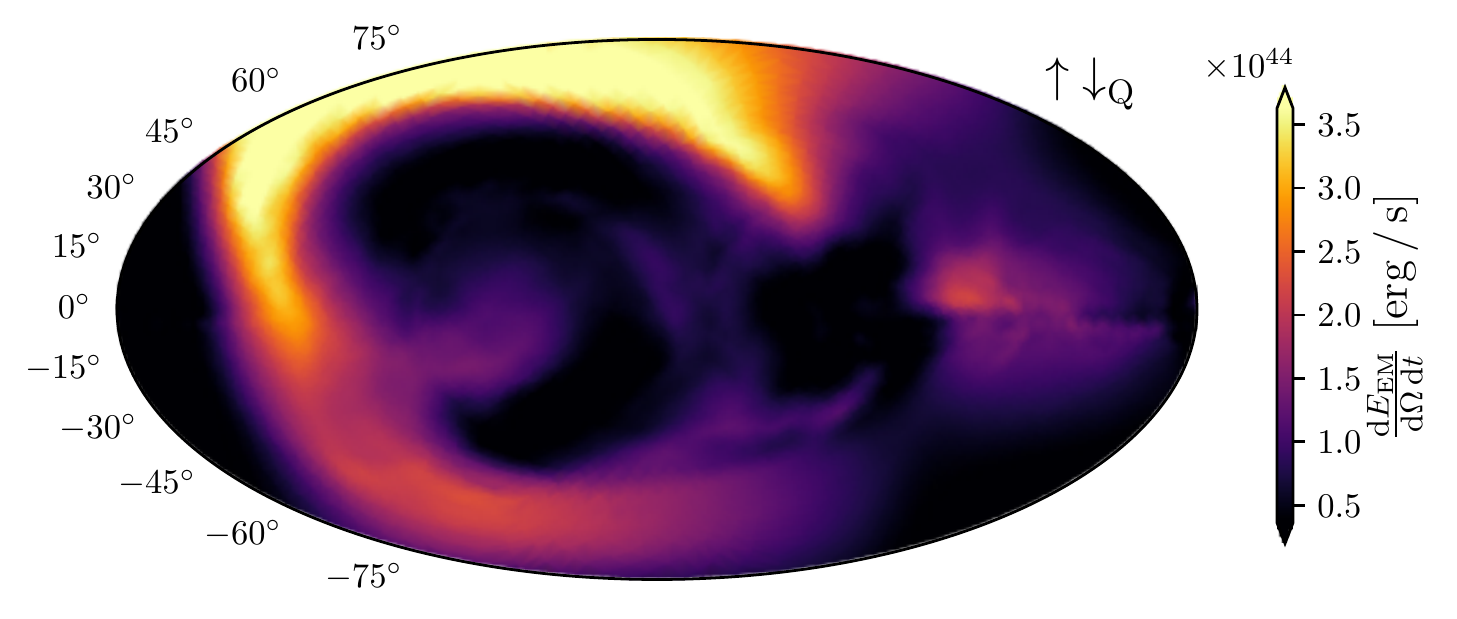}
  \caption{ Time-averaged radial Poynting flux ${{\rm d}E_{\rm EM}}/{{\rm
  d} t}$ projected onto a sphere at $r=370\, \rm km$ from the origin. 
   Same as Fig. \ref{fig:poynting2D_alignment} but showing models
   \texttt{Q}$^{\ast}\uparrow\uparrow$ and \texttt{Q}$\downarrow\downarrow$.}
  \label{fig:poynting2D_quad}
\end{figure*}
Reconnection then triggers a detachment of the magnetic bubble, with the
flare being emitted essentially along the equator (middle and right panels).
While flares kept being emitted regardless of the field strength
differences, it is apparent that a strong disparity in magnetic field strengths does
not only affect the strength of the flare but also the emission geometry.\\
In order to quantify this dependence further, we now present time-averaged
Poynting fluxes of the flares emitted in each system. This is shown in Fig.
\ref{fig:poynting2D_uneq}. The overall behavior is very similar to that
found for the different inclined models (see Fig.
\ref{fig:poynting2D_alignment}). Starting with the equal field strength
model (top left), we find that as in the anti-aligned case, two flares
are being emitted, one of them in the northern, the other in the southern
hemisphere of the binary. Both flares are equal in strength and are propagating mainly
along the polar axis. Different from the models presented in Fig.
\ref{fig:poynting2D_alignment}, here the flaring is also significantly
off-axis at the same time, with the flare covering a large angular space.
Even when decreasing the field strength of the secondary by only a factor
three (top right), the field strength of the flare decreases. Moreover,
the two flares become unequal in strength and extent, with one of them
being more strongly beamed towards the secondary. It is important to
remember that the \rev{flares are being emitted in opposite hemisphere} every half
orbit because of the relative misalignment of the dipoles.
That is, the flaring periodicity is half an orbit ($\approx 2\, \rm
ms)$, whereas the periodicity to emit a strong flare into the same
hemisphere is a full orbit ($ \approx 4\, \rm ms$). 
When further decreasing the field strength (bottom row), still two flares
are being emitted, but especially in the lowest magnetization case \texttt{U27},
the flare is weak compared to the orbital Poynting flux (left of
bottom right panel), and becomes confined to the equatorial
region.
Hence, the flaring mechanism becomes less viable in this case, as spatial
coverage and field strength both become much smaller than in the equally
magnetized case. We will quantify this behavior further in Sec.
\ref{sec:luminosity}.

\subsection{Effects of multipolar field structure}
\label{sec:quad}

\begin{figure*}
  \centering
  \includegraphics[width=0.8\textwidth]{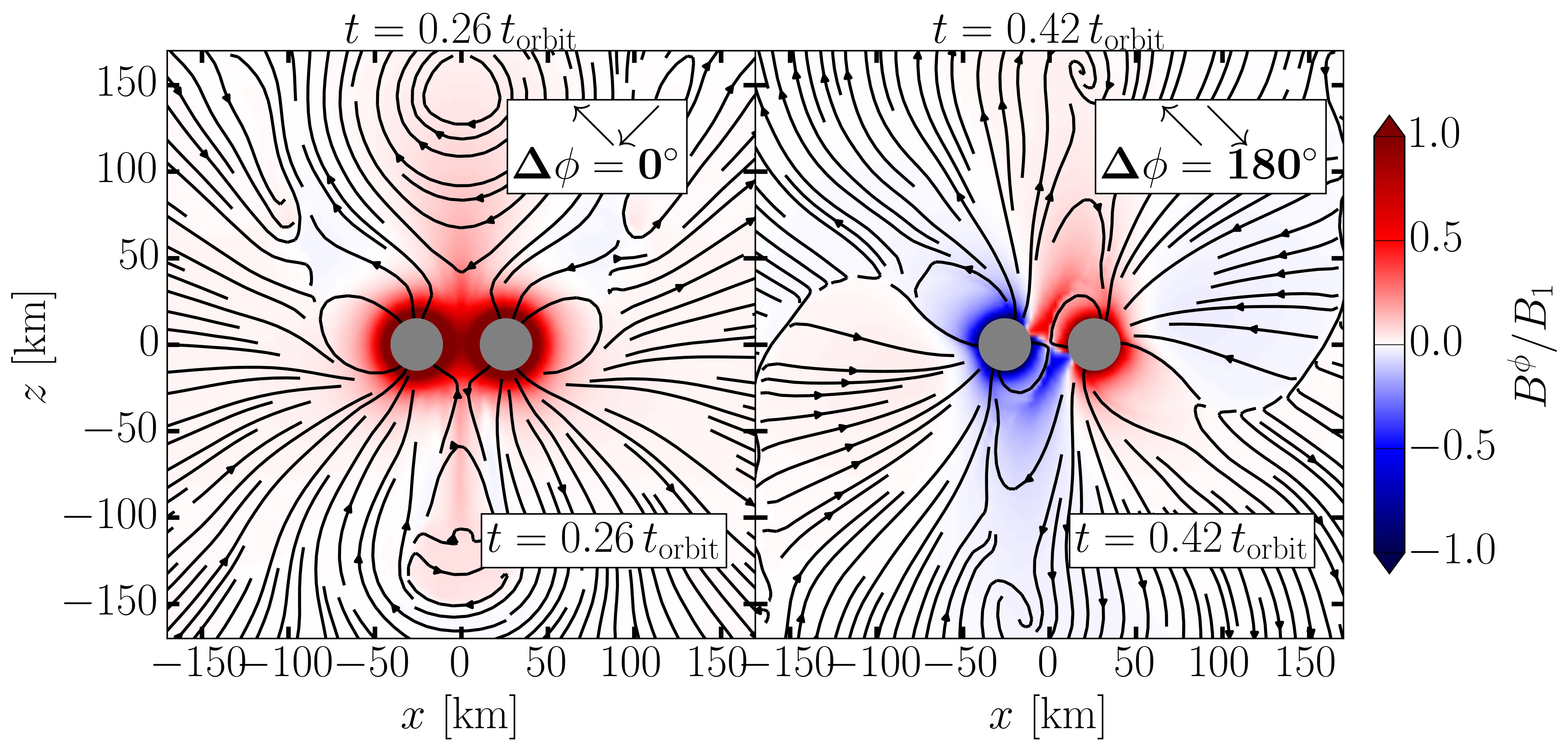}
  \caption{Impact of relative rotation $\Delta \Phi$ of two inclined dipoles (models
    \texttt{O0} and \texttt{O180}) on the final flaring state. A detailed
    description of the quantities shown in this figure is provided in the
    caption of Fig. \ref{fig:alignment}.}
  \label{fig:unequal_offset}
\end{figure*}
After showing the impact of different magnetic field inclinations and
strengths, we want to briefly comment on the effect that different magnetic
field topologies can have on the flaring process.
While the field on large scales is naturally expected to be dipolar
\cite{2004hpa..book.....L}, recent observations of PSR J0030+0451 have put forth the
possibility of having multipolar field components close to the
surface \cite{Bilous:2019knh}. Since these might be probed in the common
magnetosphere close to merger, we choose to include one fiducial model
with non-dipolar field geometry. More specifically, we chose a field structure
that globally resembles a dipolar field, while locally having an
axisymmetric quadrupolar correction \cite{Gralla:2017nbw}.
The expression for this field is provided in \eqref{eqn:A_quad}.
There, the correction is quantified in terms of a quadrupolar field strength parameter,
$\mathcal{Q}_0$. In order to investigate its impact, we consider
two configuration with $\mathcal{Q}_0$, leading either to a small or large
quadrupolar correction.\\
The interesting difference between a dipolar and a quadrupolar field
topology lies mainly in the available magnetic field lines that can form
closed flux tubes. While two aligned dipoles are not able to form connected
flux tubes, the presence of a quadrupolar part supplies additional
anti-aligned field lines that can (re-)connect. In other words, flaring in a
quadrudipole can either happen via twisting the dipolar part, which is
similar to the models considered in Sec.
\ref{sec:alignment}-\ref{sec:unequal}, or via the
quadrupolar part. To this end, we consider three models
\texttt{Q}$\uparrow\uparrow$, \texttt{Q}$^{\ast}\uparrow\uparrow$ and \texttt{Q}$\downarrow\downarrow$
with different alignments and quadrupolar strengths, see Tab.
\ref{tab:initial} for details.
As in Sec. \ref{sec:alignment}-\ref{sec:unequal}, we are going to first
describe the general flaring dynamics of this configuration and then
quantify the amount of energy in the system, that is \rev{either dissipated
or outgoing in terms of a Poynting flux.}
We first focus on the aligned configurations, \texttt{Q}$\uparrow\uparrow$,
\texttt{Q}$^{\ast}\uparrow\uparrow$, shown in the top and middle row of
Fig. \ref{fig:quadrudipole}, respectively. Looking at the northern
hemisphere of these plots, we can see that the aligned magnetic field
geometry bars the formation of connected flux tubes. Indeed no significant
build-up of out-of-plane magnetic field, $B^\varphi$, occurs. Instead,
twisting between the dipolar and quadrupolar fields of two two stars in the
southern hemisphere, leads to a built-up of twisted field lines (left
panels). However, the strength of this twist crucially depends on the
relative size of the quadrupolar correction $\mathcal{Q}_0$.
In fact, for small quadrupolar corrections, \texttt{Q}$\uparrow\uparrow$, the twist can only build up
close to the star, as the quadrupolar field lines do not extend close enough to the binary companion.
As a result, the energy stored in the twisted flux tube is small,
as is the flare. On the other hand, if the correction is significant ( \texttt{Q}${^\ast}\!\uparrow\uparrow$, middle
row), large twists can build up in the quadrupolar field, leading to strong
flaring, as in the dipole case. For anti-aligned configurations (bottom
row), twists can build up in both hemispheres, leading to the emission of
flares from the quadrupolar (left and middle panel) and dipolar sector
(right panel). The strength of the quadrupolar flare, in turn, depends on
the correction strength $\mathcal{Q}_0$.\\
We provide a more quantitative discussion in Fig.
\ref{fig:poynting2D_quad}, which shows the time averaged Poynting flux
during a flaring event projected onto a sphere at a distance of $r=370\,\rm
km$ from the origin (see Fig. \ref{fig:poynting2D_alignment} and Fig.
\ref{fig:poynting2D_uneq}). Focusing first on model
\texttt{Q}$^{\ast}\uparrow\uparrow$ (right panel), we can see that
two flares are being emitted predominantly in the polar direction.
The energy fluxes are overall similar to the other equal field strength 
models, see Fig. \ref{fig:poynting2D_alignment}.
For the misaligned case, \texttt{Q}$\downarrow\downarrow$ (right),
a strong flare is emitted, with a corresponding weaker flare in the
southern hemisphere. This flare is more inclined towards the secondary, but
consistent with the $60^\circ$ inclination of the quadrudipolar fields.\\
Overall, the flaring mechanism does not appear to be affected by 
the multipolar substructure in the magnetic field. However, the availability of the
quadrudipolar field corrections causes additional flares to be emitted. In
other words, we would expect substructure \rev{in the flaring luminosities,
  where instead of one flare, there would be secondary smaller flare
appearing}. We will discuss this point in more detail in Sec. \ref{sec:luminosity}.

\subsection{Impact of relative phase difference}
\label{sec:phase}
For systems not having magnetic moments aligned with the orbital angular momentum
another degree of freedom emerges: Relative initial offset of the two
dipoles. We investigate the impact of this, by considering two dipoles (models \texttt{O0} and \texttt{O180}), each of them being inclined by $60^\circ$ with respect to the orbital rotation axis. For one of these systems (\texttt{O180}), we include a relative phase difference of $180^\circ$ relative to the stellar rotation axis. Put differently, the initial fields are mirror images of each other.
Similar to Fig. \ref{fig:alignment}, we report the final flaring state in
terms of the out-of-plane magnetic field component $B^\phi$ (in orbital
cylindrical coordinates) in Fig. \ref{fig:unequal_offset}.
We can see that the different relative offset least to different flaring
geometries, with the flare launched roughly along the orbital axis (left panel) or completely
off-axis (right panel). Also the field strength of the twisted field
component gets modified. In fact, the electromagnetic emission
for both cases proceeds similarly, that is the flaring geometry results in
the same flares, however, their amplitudes are different, as the number of
closed field lines available is different in each case. We will discuss
this shortly in Sec. \ref{sec:luminosity} when explicitly quantifying the total amount of electromagnetic energy dissipated in the system.

\subsection{Electromagnetic energy budget}
\label{sec:luminosity}

\begin{figure}
  \centering
  \includegraphics[width=0.5\textwidth]{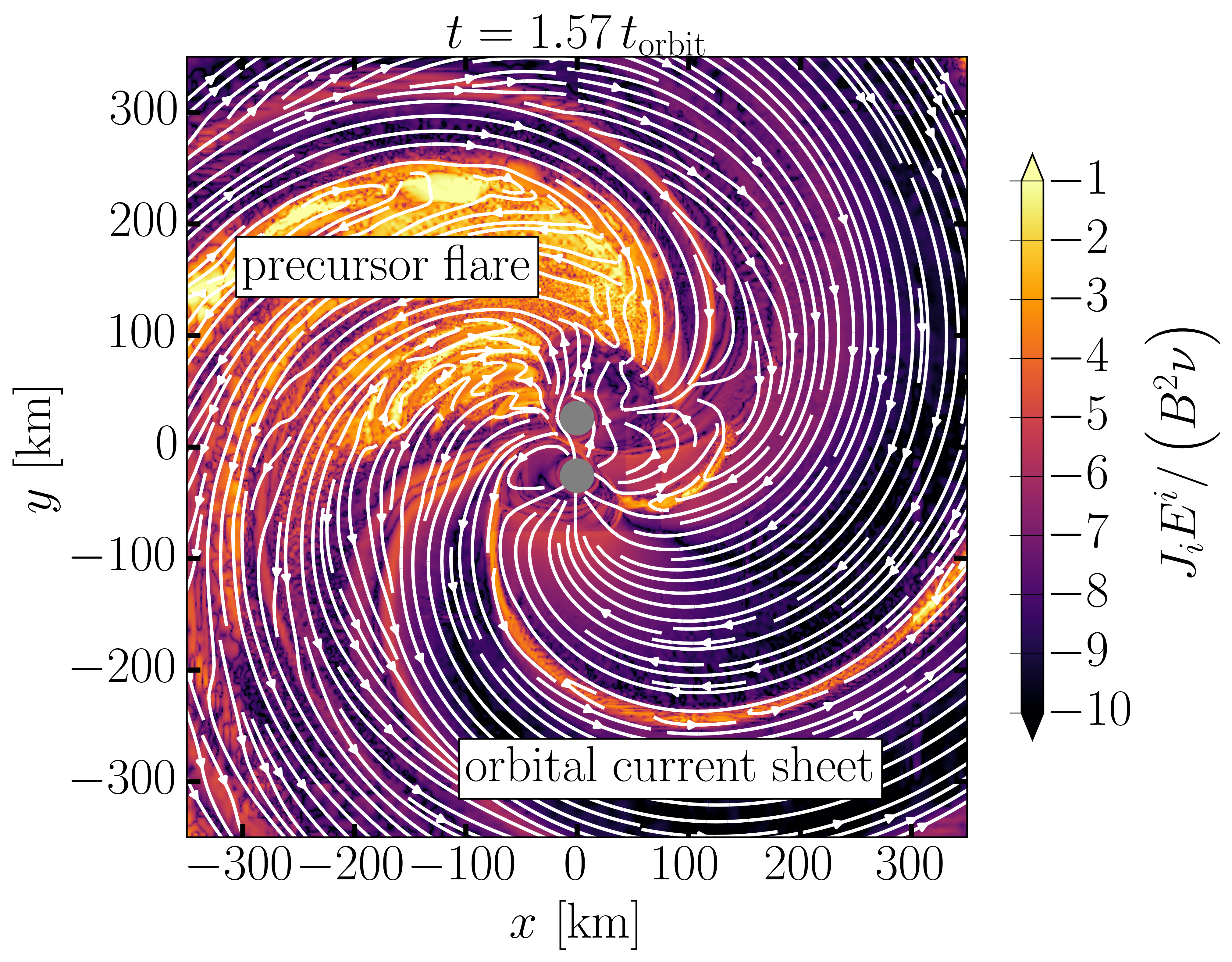}

  \caption{Dissipation rate of electromagnetic energy, $J_i E^i$,
    shown in the orbital plane of model \texttt{A30} during a flaring
  event. Highlighted are the orbital current sheet (bottom) and the dissipation
induced by the precursor flaring event (top). \rev{A large part of this is
  associated with reconnection in trailing current sheet.} The magnetic field lines are
shown as white lines projected onto the orbital plane. The normalization is set by the magnetic field strength $\sqrt{B^2}$ and the dissipation rate $\nu$.}
  \label{fig:diss_offset}
\end{figure}

Having established that the precursor flaring mechanism is viable for generic magnetic field topologies and spins in neutron star binaries, we now want to quantitatively describe
the amount of energy emission and dissipation in the system.
The latter is important for providing quantitative estimates of the amount of energy available to produce observable transients.

The ab-initio modeling of the microphysical processes powering these
transients would require accurate knowledge of the dissipation in our
simulations, e.g. in terms of the electric field in the reconnecting
current sheets, e.g. \cite{Sironi:2014jfa}.  However, the intrinsic
simplifying limitations of a force-free approach bar us from computing
first-principle  particle acceleration \rev{and dissipation} in the system.
Yet, we can still provide a meaningful upper bound on the energy available
for such processes by computing the amount of energy, which is
(numerically) dissipated in the calculations. 
Future work will be necessary to augment force-free-type simulations with
meaningful closure prescriptions for electromagnetic energy dissipation
\cite{Most:2021uck}.

\subsubsection{Quantifying dissipation}
\label{sec:analysis}

Within our simulation we can easily quantify the total amount of
electromagnetic energy $E_{\rm EM}^\mathcal{V}$ within a volume
$\mathcal{V}$ by spatially integrating Eq. \eqref{eqn:rho_EM},
\begin{align}
  E_{\rm EM}^\mathcal{V} = \int_{\mathcal{V}} \rho_{\rm EM}\, {\rm d}^3 x\,,
  \label{eqn:EV}
\end{align}
and the electromagnetic energy flux $\mathcal{S}^\mathcal{V}_{\rm EM}$ through the
boundary $\partial\mathcal{V}$ of said volume $\mathcal{V}$,
\begin{align}
  \mathcal{S}_{\rm EM}^\mathcal{V} = \oint_{\partial\mathcal{V}} \left(S_{\rm
  EM}\right)_i {\rm d} \Sigma^i\,,
  \label{eqn:EV2}
\end{align}
where ${\rm d} \Sigma^i$ is the outwards-pointing surface element.

In our simulations, dissipation will occur mainly in two places.
The first one is associated with the (inspiraling) motion of the stars.
Just as for an isolated rotating pulsar, the plasma surrounding it can only
co-rotate up until the light cylinder. This is the distance from the
rotation axis, beyond which the corotational velocity $v = \Omega \varpi$
would exceed the speed of light, i.e. $\varpi_{\rm LC} = c / \Omega$.
Beyond this point, the field lines will have to open up, separating them by
a strong current sheet \cite{Spitkovsky:2006np}. Since current sheets are the sole point of
magnetic dissipation, the continued presence of the orbital current sheet
will cause a constant amount of net dissipation, see e.g. \cite{Carrasco:2020sxg}.
This can be seen in Fig.
\ref{fig:diss_offset}, which shows the local energy dissipation rate
$J_i E^i$ in the orbital plane for model \texttt{A30}. For reference,
we estimate that the orbital
sheet is locally dissipating \rev{$\simeq 10^{29}\,\left(B/10^{12}\rm G\right)^2\, \rm
erg/\left(s\,cm^3\right) $}. During a flaring
event (top of  Fig.  \ref{fig:diss_offset}), the trailing current sheet of
the flare and its interaction with the orbital current sheet, cause
additional dissipation \rev{(in the trailing current sheet)}.  In this
specific model, \texttt{A30}, locally up to 100-times larger than in the
orbital current sheet. We emphasize that in the absence of a more realistic
\rev{Ohm's law} prescription
\cite{Most:2021uck} the precise numbers should be considered approximate.\\

In order to quantify the the amount of dissipation,
we introduce an associated luminosity $\mathcal{L}_{\rm
diss}^{\mathcal{V}}$ in a volume $\mathcal{V}$, in analogy with 
the electromagnetic energy $E_{\rm EM}^{\mathcal{V}}$ contained in that
volume.
From Eq. \eqref{eqn:EM_T_31} we find that,
\begin{align}
  \mathcal{L}_{\rm diss}^{\mathcal{V}} :=& \int_\mathcal{V} J_i E^i {\rm
  d}^3 x \nonumber\\
  =&\, \partial_t E_{\rm EM}^\mathcal{V} + \mathcal{S}_{\rm EM}^\mathcal{V}\,.
  \label{eqn:Ldiss_def}
\end{align}
Hence, we can quantify energy dissipation (which physically occurs
exclusively in current sheets) using the non-conservation of electromagnetic energy. While it would also be possible to
directly compute the dissipation rate, i.e. $J_i E^i$, from the electric current
$J_\mu$, the implicit numerical computation of $J_\mu$ based on local
violations of the force-free conditions\eqref{eqn:J} makes this
rather impractical.
In the rest of this paper, we will exclusively adopt spherical volumes
$\mathcal{V}$ with radii $r=[r_{\rm in}; r_{\rm out}]$, centered on the center of mass of the
binary. Since rotation of the two neutron stars is enforced via a boundary
condition, there is a net injection of energy from the surface of the two
stars into the computational domain. Therefore, we need to ensure that this
artificial energy injection rate is not included when evaluating
\eqref{eqn:Ldiss_def}. This can be done by choosing $r_{\rm in} > a + r_{\rm
NS}$, where $r_{\rm NS}$ is the radius of the neutron star.
In this way, the dissipation rate $\mathcal{L}_{\rm diss}$ is only computed
between two spherical shells at $r_{\rm in}$ and $r_{\rm out}$.
Using the definitions above, we may rewrite \eqref{eqn:Ldiss_def} as
\begin{align}
  \mathcal{L}_{\rm diss} = \mathcal{S}_{\rm EM}^{r_{\rm out}} -
  \mathcal{S}_{\rm EM}^{r_{\rm in}} + \frac{ {\rm d}}{ {\rm d}t} \left(
  E_{\rm EM}^{r_{\rm out}} - E_{\rm EM}^{r_{\rm in}} \right)\,.
  \label{eqn:Ldiss_eval}
\end{align}
For the remainder of this work we choose $r_{\rm in}=2.3\, a$ and $r_{\rm
out}=12.3\,a$, where $a$ is the orbital separation, see Tab. \ref{tab:initial}. We have ensured that the flaring happens both inside the
shell and that the choice of outer radius has negligible effect on the
extraction of the Poynting flux and energy dissipation rate.


\subsubsection{Energy budget of a precursor flare}

\begin{figure}
  \centering
  \includegraphics[width=0.4\textwidth]{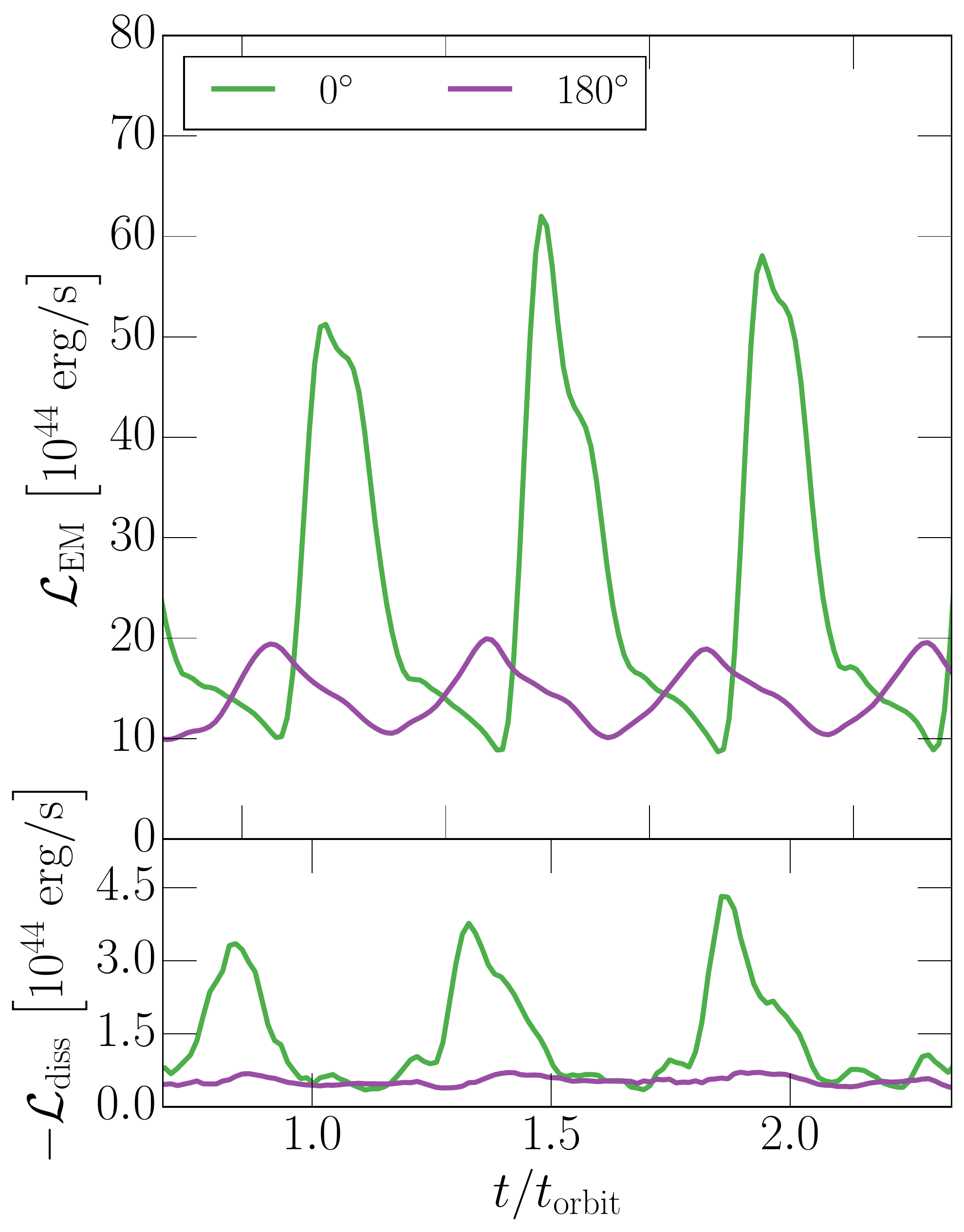}
  \caption{Electromagnetic luminosity $\mathcal{L}_{\rm EM}$ of the precursor flares
  for different relative rotation $\Delta \Phi$ of the two dipoles (models \texttt{O0}
  and \texttt{O180}), in an equally magnetized
  neutron star binary. (Top) Positive values refer to outgoing Poynting
fluxes computed on a sphere with radius $r=12.3 a$, where $a$ denotes the orbital separation. (Bottom) The dissipative
luminosity $\mathcal{L}_{\rm diss}$ refers 
to the energy dissipated in current sheets within a spherical shell between
$r_{\rm in} = 2.3\, a$ and $r_{\rm out} = 12.3\, a$. The times $t$ are expressed relative to the orbital period $t_{\rm orbit}$.
}
  \label{fig:offset_poynting}
\end{figure}

Following our discussion of how to quantify dissipation in current
sheets, Sec. \ref{sec:analysis}, we now compute the energy budget of the
flare in terms of dissipated energy and outgoing Poynting flux.
Before presenting a comparison of the flares obtained for different
magnetic field inclinations (Sec. \ref{sec:alignment}), magnetizations (Sec.
\ref{sec:unequal}) and topologies (Sec. \ref{sec:quad}) , we illustrate the analysis using models \texttt{O0} and
\texttt{O180} presented in Sec. \ref{sec:phase}.
The resulting luminosities are shown in Fig. \ref{fig:offset_poynting}.
In both cases we find a periodic signal in the Poynting flux (top curves). 
The periodicity of the flares corresponds to half of the orbital time
scale, as expected. The initial offset of the dipoles translates to a
quarter orbital time shift in the flares, the strength of which changes by
about a factor three between the two configurations.
We point out that when (gravitational wave) radiation backreaction was
included, the strength of the Poynting flux would
increase with time, whereas the period would shrink, as the two stars
inspiral.
Looking at the dissipation (bottom panel), we find clear periodic dissipation associated
with the flares in the \texttt{O0} model with $0^\circ$ relative phase
shift between the dipoles. In the \texttt{O180} model, however, the
dissipation in trailing current sheet is strongly suppressed. Indeed, the
almost constant level of dissipation coincides with the lower dissipation
limit of the  \texttt{O0} model. This is a strong indication that
dissipation is dominated by the contribution from the orbital current sheet that is constantly sourced by the orbital motion of the stars, see Fig.
\ref{fig:diss_offset}. \\
While the emission associated with the direct emission of a magnetic bubble
in the flare will always be present, transients associated with dissipation
in the trailing current sheet \cite{Philippov:2019qud}, might
strongly depend on the relative phase offsets of the two magnetic fields.
For surface magnetic field strengths of $\simeq 10^{12}\, \rm G$, the luminosities of are always of the order
$10^{44}-10^{45} \rm erg/s$ with dissipation being at least an order of
magnitude smaller. This is consistent with our earlier findings of the
flaring process \cite{Most:2020ami}.

\subsubsection{Dependence on magnetic field topology}

\begin{figure*}
  \centering
  \includegraphics[width=\textwidth]{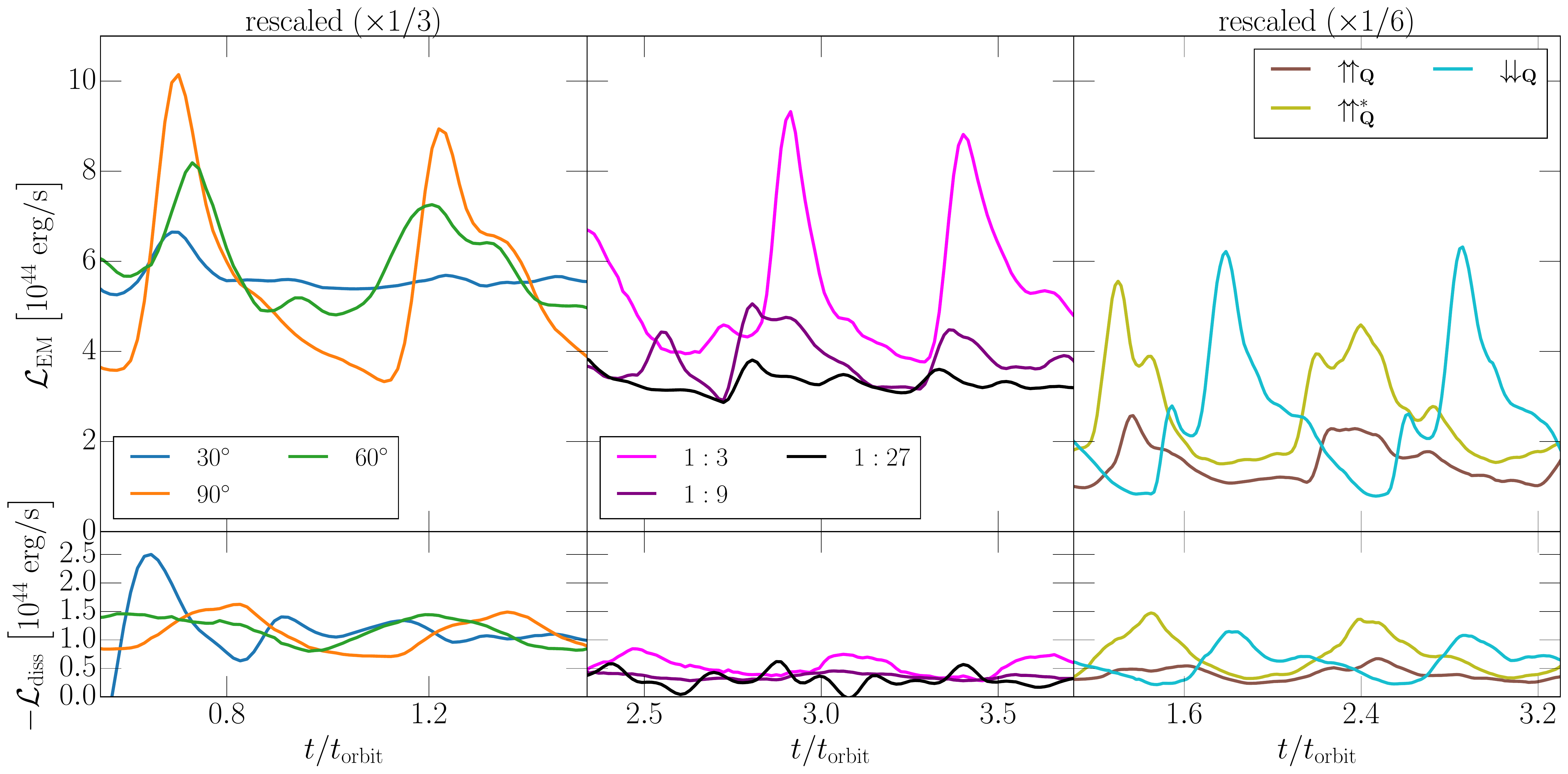}
  \caption{Average Poynting fluxes for different inclinations (left panel),
  magnetization ratios (middle panel) and quadrudipolar field configuration
  (right panel). See caption of Fig. \ref{fig:offset_poynting} for a
description of the quantities shown.}
  \label{fig:poynting_all}
\end{figure*}

Following our detailed description of the flaring process and its
energetics, we are finally in a position to compare the energetics of the
different models considered in this work. More specifically, we are
interested in understanding how the total luminosity of the flare and the
dissipation it induces scale with the inclination of the magnetic field,
and the field strength of its companion. To this end, we show the
energetics in Fig. \ref{fig:poynting_all}, in a similar fashion to Fig.
\ref{fig:offset_poynting}. Starting with the case of relative magnetic field
inclination but equal magnetic moments, models \texttt{A0}--\texttt{A90},
we find that the luminosity of the flares vary within a factor $L_{\rm
EM}/L_{\rm EM}^{0^\circ}\simeq 2$ relative to the aligned system.
Looking at the lower left panel of Fig.
\ref{fig:poynting_all}, we find that the dissipated energy rate is about a
factor $5$ smaller in all cases. Despite the clear dependence of the
luminosity of the flare on the relative size of the magnetic moments of the
two neutron stars, the dissipation rate in the trailing current sheet is
roughly the same in either case. This is similar in behavior and magnitude
to the cases with different inclination angle. This strongly indicates the the energy
dissipated in current sheet might be largely independent of the choice of 
magnetic field topology.
For the case of the quadrudipole topology (right panels) the situation is
similar. Small
variations in the energy dissipation rate exist, but for equal
magnetization the overall dissipation rate is roughly the same.
The Poynting fluxes of all three cases show mild variations, with repeated
flaring activity. The strength of the flares directly correlates with the
amount of flux tubes available for the flaring process. Indeed, for the
smallest quadrudipole configuration in the aligned cases
($\uparrow\uparrow_Q$) only a tiny fraction of field lines from the
quadrupolar part can be twisted, see also Fig. \ref{fig:quadrudipole}.
This increases with stronger quadrupole component ($\uparrow\uparrow_Q$).
Indeed, for the anti-aligned cases both the quadrupolar and dipolar part can
be twisted. This results in a double peak sub-structure, with the strong flares
corresponding to the dipolar, the weaker once to the quadrupolar part.
The overall luminosity only varies within a factor 2 between the different
configurations.

This leads us to conclude that both the dissipation, as well as the flaring
luminosities are largely invariant under changes of the magnetic field
topology. Flaring will only start to cease if the system becomes very
unequally magnetized, e.g. for model \texttt{U27}.

\subsubsection{How many flares do we expect to be observable?}

Now that we have quantified the energetics of the flaring process, we want
to estimate how many flares we would expect to be in principle observable.
While a detailed discussion of the microphysical emission process is beyond
the scope of this paper and will be addressed in a forthcoming work, we can
already place limits based on the total electromagnetic luminosity
available in the process. This will provide an estimate for the upper bound for the
available electromagnetic energy radiated per flaring event. \rev{We will
separately consider radio and X-ray emission.}\\
To this end we will make the following assumptions.
First, for slowly rotating or irrotational systems, a flare will be
launched after a $180^\circ$ twist. This will happen roughly twice per orbital
period. The exact expression for arbitrary binary configurations
can be found in Ref. \cite{Cherkis:2021vto}.
We recall that that the orbital frequency $f$ and the frequency of the
gravitational wave signal $f_{\rm GW}$ are related by, $f_{\rm GW} = 2 f$.
Hence, the number of gravitational wave cycles and of flares coincides, and
we can write
\begin{align}
  N_{\rm flares} = \int f_{\rm GW} {\rm d}t =  \int_{f_{\rm min}}^{f_{\rm max}} \frac{\dot{f}_{\rm
  GW}}{f_{\rm GW}} {\rm d} f_{\rm GW}\,,
  \label{eqn:Ncyl1}
\end{align}
where $f_{\rm min}$ and $f_{\rm max}$ are initial and final frequency
bounding the interval of the inspiral with $N_{\rm flares}$
gravitational-wave cycles. 
Assuming that $f_{\rm max} \gg f_{\rm min}$, one can integrate 
\eqref{eqn:Ncyl1} to yield
\cite{maggiore2008gravitational},
\begin{align}
  N_{\rm flares} = \frac{1}{32 \pi^{8/3}} \left( \frac{G
    \mathcal{M}}{c^3} f_{\rm min}\right)^{-5/3} \,.
  \label{eqn:Ncyl2}
\end{align}
Here, we have introduced Newton's constant $G$, the speed of light $c$ and
the chirp mass $\mathcal{M} =  \left[ q/\left( 1+q \right)^2 \right]^{3/5}
M$, which is determined by the binary masses $m_1$ and $m_2$ in the form of
the total mass $M= m_1+m_2$ and the mass ratio $q=m_2/m_1$.

The initial frequency $f_{\rm min}$ is determined by the time, when the
flaring luminosity becomes large enough to (in principle) be observable.
In good agreement with the present simulations, we previously found that \cite{Most:2020ami},
\begin{align}
  \mathcal{L}_{\rm EM} = {\mathcal{L}}_{0}\left(
  \frac{a_0}{a}\right)^{7/2}\,,
  \label{eqn:Lscaling}
\end{align}
where $a_0=45\,\rm km$ and 
\begin{align}
\mathcal{L}_0 &= 7\times 10^{44}\,\sigma \left( \frac{B_1}{10^{12}\, \rm G}\right)^2\,,
\rm erg/s\,, \label{eqn:L0_this_study}
\end{align} 
for the luminosity of the expanding magnetic bubble, for which the efficiency $\sigma =1$.
Here, we have introduced the efficiency $\sigma = 0.1$ for X-rays associated with the dissipation in the current sheet, and $\sigma=10^{-5}-10^{-3}$ for radio transients directly associated with the outgoing magnetic bubble.
This is true for both the synchrotron maser mechanism at the shock launched by the bubble in the binary wind \cite{Plotnikov:2019zuz,Beloborodov:2020ylo}, 
as well as the emission from plasmoid mergers \cite{Lyubarsky:2018vrk,Philippov:2019qud} expected from the collision of the flaring bubble and the magnetospheric current sheet \cite{Lyubarsky:2020qbp}. 
Other characteristics of the observed signal, 
such as characteristic frequencies and burst duration, will be studied in a forthcoming work.
\\

From Kepler's law, we obtain $16 \pi^2 f_{\rm GW}^2 = G M/\varpi^3 $. 
Hence we can relate,
\begin{align}
  f_{\rm min} = \frac{1}{16 \pi^2} \left( G \mathcal{M} \left[
  \frac{q}{\left( 1+q \right)^2}\right]^{3/5} \right)^{1/2} \varpi_0^{-3/2}
  {\mathcal{L}}_{0}^{-3/7} {\mathcal{L}}_{\rm EM}^{3/7}\,.
  \label{eqn:fmin_L}
\end{align}
Combining \eqref{eqn:fmin_L} with \eqref{eqn:Ncyl2}, we obtain the final
scaling
\begin{align}
  &N_{\rm flares} = \frac{\left(8\pi\right)^{2/3}}{2^{1/3}} \varpi_0^{5/2}
  \mathcal{L}_0^{5/7}\, 
   \left[
  \frac{\left( 1+q \right)^2}{q} \right]^{1/2} \nonumber\\
  &\phantom{
  N_{\rm flares} = \frac{\left(8\pi\right)^{2/3}}{2^{1/3}} \varpi_0^{5/2}
  \mathcal{L}_0^{5/7}\,}
   \left( \frac{G^2
      \mathcal{M}^2}{c^3} \right)^{-5/3} \mathcal{L}_{\rm min}^{-5/7} \,,
\end{align}
where $\mathcal{L}_{\rm min}$ is the minimum luminosity required for a detection.
Assuming a near-equal mass binary, $q\approx 1$, with canonical neutron
star masses $m_1=m_2=1.4\, M_\odot$, we find 
\begin{align}
     \phantom{ N_{\rm flares} }\simeq 17   \left(\frac{B}{10^{11}\, \rm G} \right)^{10/7}\, \left(
  \frac{1.2 M_\odot}{\mathcal{M}} \right)^{5/2}\,& \left(\frac{10^{42}\, \rm
erg/s}{\mathcal{L}_{\rm min}}\right)^{5/7} \nonumber\\ &\left(\frac{\sigma}{10^{-4}}\right)^{5/7}
      \,.
  \label{eqn:Ncyl2_apl}
\end{align}
We can see that the number of potentially observable flares will strongly
depend on the magnetic field strength and the emission mechanism.
More specifically \eqref{eqn:Ncyl2} implies, that for magnetic field
strengths $\ll 10^{11}\,\rm G$ no
flares will be observable. Effects such as unequal magnetization would
further decrease the number of flares, as the luminosity can be suppressed
by a factor of a few, see Fig. \ref{fig:poynting_all}.
Furthermore, the above estimate does not account for anisotropies in the
emission, e.g. beaming effects. This would raise the required 
luminosity $\mathcal{L}_{\rm min}$ further suppressing the number of
potentially observable flares.

\section{Conclusions}
\label{sec:conclusions}

In this work, we have investigated the impact of different magnetic field
topologies on the precursor flaring mechanism in coalescing neutron star
binaries \cite{Most:2020ami}. The interaction of oppositely directed
magnetic field configurations in neutron star binaries can lead to the
built-up of twists in the common magnetosphere, that will ultimately
culminate in the release of a powerful electromagnetic flare, not unlike
in a coronal-mass ejection event in the Sun\cite{2011LRSP....8....1C}.

Focusing on different magnetic field topologies, we have investigated the
viability of producing precursor flares in the late inspiral of neutron
star binaries. More specifically, we have presented a new set of 13
simulations investigating different magnetic field inclinations,
magnetization ratios and quadrudipolar field topologies.
One of our main findings is, that the emission of flares will always happen
unless one of the two neutron star is significantly less magnetized than
the other. In particular, we find a strong suppression of the luminosity of
a flare of the field strength differ by ratios much greater than 1:10.
Furthermore, we find that the inclusion of higher order multipolar field
structure leads to secondary flaring events, when, e.g., the quadrupolar
part gets twisted and flares in addition to the twist on the dipolar part.
While the production mechanism of flares itself is interesting, their interaction
with the surrounding medium might lead to the production of radio 
transients (e.g., \cite{Metzger:2016mqu}). We have also
quantified the amount of dissipation induced by the presence of the flare. 
In an earlier work \cite{Most:2020ami}, we had shown that the trailing current sheets in this
process lead to an order-of-magnitude lower luminosity, that is able to
power (largely) X-ray transients \cite{Beloborodov:2020ylo}.
Interestingly, we find that the amount of dissipation is largely
independent of the flaring geometry. We do find (sometimes small) periodic
enhancement in the energy dissipation rate corresponding with individual
flaring events. Yet, the amount of dissipation varies at most by a factor
of a few \rev{compared to the dissipation in the orbital current sheet.
We remark, that if the field strength in the flare becomes too weak, i.e
$B_2 \ll 10^{-2} B_1$, dissipation will be dominated completely by that of the
orbital current sheet.}

While our results strongly hint at the presence of electromagnetic flares
in the late inspiral of neutron star binaries, we caution that these
systems contain inherent uncertainties. Firstly, our study assumes the
presence of a force-free pair-plasma in the common magnetosphere
\cite{Goldreich:1969sb}. While this might be a reasonable assumption in the
late stage of the binary, where the orbital motion can induce strong
electric fields \cite{Lyutikov:2018nti,Wada:2020kha}, it will likely
require the presence of sufficiently large magnetic fields \cite{1992MNRAS.255...61M}.
Finally, systems with fully aligned magnetic moments (and only dipole
fields) will also not be able to sustain any flaring activity.

Apart from those concerns, our parameter coverage is by no means
exhaustive. Although we have only focused on a single orbital separation
representative of the late inspiral, we have shown earlier, that the
flaring strength strongly correlates with the orbital separation
\cite{Most:2020ami}. However, changes in the
separation will only affect the orbital velocity and, hence, for
most realistic binaries with vanishing stellar spins
\cite{Bildsten:1992my}, only the periodicity of the flares.
On the other hand, extreme spins might leave strong imprints on the
post-merger dynamics (e.g.,
\cite{Dietrich:2016lyp,Ruiz:2019ezy,East:2019lbk,Most:2019pac,Chaurasia:2020ntk,Papenfort:2022ywx})
and life-time of the system \cite{Tootle:2021umi,Papenfort:2022ywx}.
In terms of the precursor flaring mechanism they would largely affect the
periodicity, since the effective rotation frequency in the corotating frame 
will be $\Omega-\omega$ \cite{Lai:2012qe}, see also \cite{Cherkis:2021vto}
for a complete discussion and more detailed expression for general spins
and magnetic field alignments. 
We note that in the absence of tidal locking \cite{Bildsten:1992my}, the
stars are always expected to have an effective twist rate relative to the
orbital motion. We intend to provide such a thorough investigation of \rev{spin
effects} in a future study.

Finally, the modeling of the plasma physics in this work can, at best, be
considered rudimentary.  Meaningful simulations require the use of more
accurate Ohm's law closures for the electric field \cite{Bessho:2005zz}.
Future work using novel approaches to relativistic two-fluid systems will
be required to improve the global modeling of these transient phenomena
\cite{Most:2021uck}.

\section*{Acknowledgments}
The authors are grateful for discussions with S. Cherkis, S. de Mink,
G. Hallinan, D. Lai, M. Lyutikov and K. Mooley.
ERM gratefully acknowledges support from a joint fellowship
at the Princeton Center for Theoretical Science, the Princeton Gravity
Initiative and the Institute for Advanced Study. The simulations were
performed on the NSF Frontera supercomputer under grants AST20008 and
AST21006. AP acknowledges support by the National Science Foundation under grant No. AST-1909458. We acknowledge the use of the following software packages: AMReX \cite{amrex}, matplotlib \cite{Hunter:2007}, numpy
\cite{harris2020array} and scipy \cite{2020SciPy-NMeth}. 


\bibliography{inspire,non_inspire}

\end{document}